\documentclass[prb,final,twocolumn,showpacs,fixfloat,nofootinbib,superscriptaddress]{revtex4-1}
\usepackage{amsmath}
\usepackage{natbib}
\usepackage[pdftex]{graphicx}
\usepackage{epstopdf}
\usepackage{tabularx}
\usepackage[colorlinks,bookmarks=false,citecolor=blue,linkcolor=red,urlcolor=blue]{hyperref}
\usepackage[normalem]{ulem}

\def\be {\begin{equation}}
\def\ee {\end{equation}}
\def\etal{{\it et~al.}}
\def\gsim{\mathrel{\rlap{\lower3pt\hbox{\hskip0pt$\sim$}}\raise1pt\hbox{$>$}}}
\def\lsim{\mathrel{\rlap{\lower4pt\hbox{\hskip1pt$\sim$}}\raise1pt\hbox{$<$}}}

\begin{document}

\title{Zero-temperature Monte Carlo study of the non-coplanar phase of the classical bilinear-biquadratic Heisenberg model on the triangular lattice}
\author{Sandro~Wenzel}
\date{\today}
\affiliation{Institute of Theoretical Physics, Ecole Polytechnique
F\'ed\'erale de Lausanne, CH-1015 Lausanne, Switzerland}
\author{Sergey~E.~ Korshunov}
\affiliation{L.~D. Landau Institute for Theoretical Physics, Russian Academy of Sciences, 142432
Chernogolovka, Russia}
\author{Karlo Penc}
\affiliation{
Institute for Solid State Physics and Optics, Wigner Research 
Centre for Physics, Hungarian Academy of Sciences, P.O.B. 49, 
H-1525 Budapest, Hungary}
\author{Fr\'ed\'eric Mila}
\affiliation{Institute of Theoretical Physics, Ecole Polytechnique
F\'ed\'erale de Lausanne, CH-1015 Lausanne, Switzerland}


\begin{abstract}
We investigate the ground-state properties of the highly degenerate 
non-coplanar phase of the classical bilinear-biquadratic Heisenberg 
model on the triangular lattice with Monte Carlo simulations. For that 
purpose, we introduce an Ising pseudospin representation of the ground 
states, and we use a simple Metropolis algorithm with local updates, 
as well as a powerful cluster algorithm. At sizes that can be sampled 
with local updates, the presence of long-range order is surprisingly 
combined with an algebraic decay of correlations and the complete 
disordering of the chirality. It is only thanks to the investigation 
of unusually large systems (containing $\sim 10^8$ spins) with cluster 
updates that the true asymptotic regime can be reached and that the 
system can be proven to consist of equivalent (i.e., equally ordered) 
sublattices. These large-scale simulations also demonstrate that the 
scalar chirality exhibits long-range order at zero temperature, 
implying that the system has to undergo a finite-temperature phase 
transition. Finally, we show that the average distance in the order 
parameter space, which has the structure of an infinite Cayley tree, 
remains remarkably small between any pair of points, even in the limit 
when the real space distance between them tends to infinity. 
\end{abstract}

\maketitle

\section{\label{sec:introduction}Introduction}
It is hardly possible to overestimate the role played by the Heisenberg model with interactions of the type ${\bf S}_i\cdot{\bf S}_j$ between nearest
neighbors in magnetism and more generally in statistical physics. However, when
dealing with quantum spins $S$ larger than 1/2, the most general rotationally invariant interaction is actually
a polynomial of degree $2S$ in  ${\bf S}_i\cdot{\bf S}_j$. These models play a very important role in
one dimension because, for specific polynomials, exact results have been derived, either because the model
is integrable,\cite{sutherland,takhtajan,babujian} or because the ground state can be constructed explicitly.\cite{AKLT}
For spin $1$,  the corresponding Hamiltonian is  the
bilinear-biquadratic Heisenberg model defined by
\begin{equation}
\label{eqn:Hamiltonian}
  \mathcal{H} =  J_1 \sum_{(ij)} {\bf S}_{i} \cdot {\bf S}_{j}
  + J_2 \sum_{(ij)} \left({\bf S}_{i} \cdot {\bf S}_{j}\right)^2\;.
\end{equation}
The coupling constants $J_1$ and $J_2$ are conveniently parametrized by $J_1=J\cos \theta$ and
$J_2=J\sin \theta$ with $\theta\in[-\pi,\pi)$.
A lot of attention has been focused recently on studying the same model
on the triangular lattice, following the observation that
the compound NiGaS$_4$ does not
possess any trace of standard periodic magnetic ordering at low
temperatures.\cite{Nakatsuji2005}
 Accordingly, it has been suggested that the ground state is
either a completely disordered spin-liquid or a
spin-nematic phase.  This suggestion has been supported by several theoretical investigations\cite{Harada2002,Tsunetsugu2006,Bhattarchajee2006,Laeuchli2006}
which have demonstrated that the spin $1$ bilinear-biquadratic Heisenberg model \eqref{eqn:Hamiltonian} on the triangular can indeed stabilize ferro- or antiferroquadrupolar order. Similar results have been obtained on the square lattice.  \cite{Toth2012}

It has been shown by Kawamura and Yamamoto \cite{Kawamura2007} that in the case of the triangular lattice the model defined by Eq.~\eqref{eqn:Hamiltonian} has a rich phase diagram with various competing phases even in the classical limit when spins ${\bf S}$ are classical vectors of unit length. After analyzing which spin configurations minimize the energy of a single triangular plaquette for different values of the parameter $\theta$, these authors have proposed a zero-temperature phase diagram with the notable presence of four different phases: a ferromagnetic phase(I), a spin-nematic phase(II), a three-sublattice antiferromagnetic ($120^\circ$) phase (III), and a non-coplanar phase (IV).

While  phases I-III are rather well understood on the basis of the results of Ref.~\onlinecite{Kawamura2007} (which includes also numerical simulations), a thorough discussion of the non-coplanar phase (IV) was only recently given in Ref.~\onlinecite{Korshunov2012} by three of the present authors. In that paper, the family of possible ground states has been systematically analyzed, and it has been shown that they can be represented in terms of domain wall configurations, giving rise to a loop model. It was further shown that the non-coplanar phase has an extensive ground-state degeneracy, which makes the question of whether at zero temperature the system is magnetically ordered, critical, or disordered a non-trivial one. By constructing a classification of the non-coplanar ground states in terms of an infinite Cayley tree of degree 3, and by developing an analogy to an $O(3)$ model on a honeycomb lattice that has the same ground-state manifold, the authors of Ref.~\onlinecite{Korshunov2012} have argued that these states are, except for two special values of the parameter $\theta$, magnetic states with a true long-range order in terms of spin orientation.

The main goal of this paper is to critically address this
claim with an unbiased numerical approach based on a Monte Carlo (MC)
method capable of efficiently sampling non-coplanar ground-state
configurations. Another important motivation is to investigate whether the non-coplanar ground states are of chiral nature, since this question is beyond the scope of the analysis of Refs.~\onlinecite{Kawamura2007} and \onlinecite{Korshunov2012}.
As we shall see, this task turns out to be rather involved, with the need to
go to very large systems with roughly $10^8$ spins in order to reach a reliable conclusion.

The paper is organized as follows. Sec.~\ref{sec:mapping} reviews the
classification of non-coplanar ground states of the bilinear-biquadratic model on the triangular lattice in terms of the zero-energy
domain walls introduced in Ref.~\onlinecite{Korshunov2012}.
In this section, we also show that these ground states can be mapped
onto the states of an Ising-like model (pseudospin) defined on the same triangular lattice, which facilitates the numerical analysis of the family of ground states.  Section~\ref{sec:MTmetro} presents Monte Carlo
results obtained for this Ising-like model
using a local update technique on medium system sizes.
While those numerical results are already in favor of magnetic ordering, they do not allow one to understand the precise nature of ordering and
suggest the necessity to go to much larger length scales for clear conclusions.
In Sec.~\ref{sec:MTcluster} we introduce a non-local update mechanism capable of efficiently sampling the ground-state manifold up to very large system sizes and allowing to draw unambiguous conclusions regarding the
nature of the phase.  Section~\ref{sec:MTcluster:results} is devoted
to a presentation of those numerical results, and in
Sec.~\ref{sec:conclusions} the conclusions and a summary are given.

\section{\label{sec:mapping}Ground-state manifold and the pseudospin representation}
\subsection{Ground-state manifold\label{sec:loop model}}

In the classical model defined by Hamiltonian  Eq.~(\ref{eqn:Hamiltonian}), the energy of the interaction of two neighboring spins is given by
\begin{equation}
E(\Phi)=J_1 \cos \Phi + J_2 \cos^2 \Phi\,,
\end{equation}
where $\Phi$ is the angle between them. For $|J_1|\leq 2J_2$ this energy is minimized when $\Phi = \arccos\left(-{J_1}/{2J_2}\right)$. Accordingly, in the window $-2J_2<J_1<J_2$, the angle $\Phi$ belongs to the interval $0<\Phi<2\pi/3$, which does not lead to frustration on triangular plaquettes. For such values of $\Phi$, the energies of all bonds can be simultaneously minimized on the whole triangular
lattice, leading to a highly degenerate manifold of non-coplanar ground states. Our aim is to study the nature of ordering in the manifold of such states.

\begin{figure}
\centering
\includegraphics[width=0.8\columnwidth]{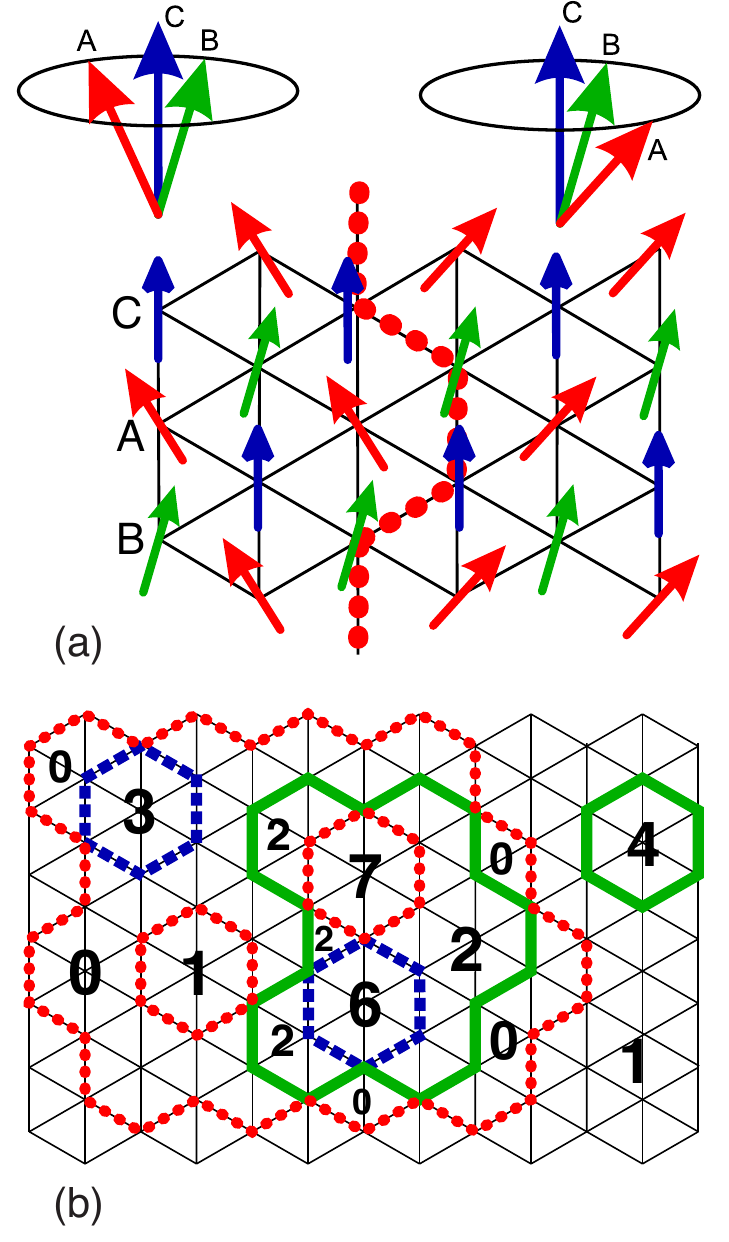}
\caption{\label{fig:loops}(Color online) (a) Part of a triangular lattice
  with sublattices $A$, $B$, $C$ and a simple domain wall of type $A$
  (on links $BC$) separating two domains having spins as represented by the two spin triads.
  The spins on sublattice $A$
  differ in the two domains by the reflection with respect to the
  plane formed by the spins on the other two sublattices, which on
  both sides of the wall have the same orientation.
  (b) Visualization of a more complex
  domain wall configuration (without showing spins) giving rise to a set of domains of different regular states labelled (in an arbitrary way) by numbers from $0$ to $7$.} 
\end{figure}

For the classification  of all possible ground states, it is convenient to partition the triangular lattice into three equivalent triangular sublattices which  are labelled $A$, $B$, and $C$ below.
When the angles between neighboring spins are all equal to $\Phi$, neighboring spins on the same sublattice have either to be parallel to each other, or to make an angle $\Phi_2=2\arccos\frac{\cos\Phi}{\cos(\Phi/2)}$.
It is convenient to say that in the latter case the spins
are separated by a zero-energy domain wall,\cite{Korshunov2012} see Fig.~\ref{fig:loops}(a).
These domain walls live on the bonds of the honeycomb lattice composed by the sites of the remaining two sublattices (e.g. $B$ and $C$ if we consider domain walls for the spins on sublattice $A$), so that the bonds of the triangular lattice are partitioned into three distinct and non-overlapping interpenetrating sets of bonds of three honeycomb lattices (consequently, we use three different colors to draw them, e.g., blue for walls passing through $A$ and $B$ sites, red for those passing through $B$ and $C$ sites and green for those passing through $C$ and $A$ sites (see Fig.~\ref{fig:loops}).

With this convention, any possible ground state can be characterized (or constructed) by specifying on some plaquette three spins that form angles $\Phi$  with each other \emph{plus} a configuration of domain walls. \cite{Korshunov2012}
In this language, the simplest ground states in which all spins on the same sublattice have the same direction (the \emph{regular states} with three-sublattice structure) correspond to the absence of domain walls.
Any other ground state can be constructed from a regular state by introducing an appropriate set of domain walls on the aforementioned bonds of the honeycomb lattices.

For generic $\Phi$, the domain walls can end only at the boundary (or have to be closed) and cannot cross each other. Generic $\Phi$ here means
\begin{equation}                                           \label{eq:Phineq}
\Phi\neq\pi/2,\,\arccos(-1/3)\,.
\end{equation}
For these two specific values of $\Phi$ the local rules allow also for some other domain walls configurations which transform the model into an exactly solvable one. \cite{Korshunov2012} The analysis below always implies the fulfillment of condition (\ref{eq:Phineq}).

The domain walls divide the system into \emph{domains} of different
regular states, see Fig.~\ref{fig:loops}(b) for an example. Upon
crossing a domain wall passing through sites $B$ and $C$, one reaches another
domain in which the spins on sublattice $A$ are reflected with respect to
the plane formed by the spins on two other sublattices, see
Fig.~\ref{fig:loops}(a). Similar rules hold for the other two types of domain
walls. Summarizing, all ground states may be generated by systematically
enumerating all possible domain wall configurations that exist on a lattice of a given size and by specifying which regular state is realized in one of the
domains.

\subsection{Pseudospin representation}

In order to study the problem of ordering in non-coplanar states with
Monte Carlo simulations, it is necessary to efficiently sample the space of
allowed domain wall configurations. The simplest way to
enumerate such configurations consists in introducing Ising-like (binary) variables
$\sigma_i=\pm 1$ (pseudospins) defined on the sites of the same triangular lattice in such a way that each segment of a domain wall separates two pseudospins of opposite signs which are next-to-nearest neighbors of each other. It can be easily checked that the application of this rule does not lead to any contradiction for all allowed configurations of domain walls passing
through a given site.

The existence of such a mapping allows one to replace the analysis of the loop model described in Sec.~\ref{sec:loop model} by that of the Ising-like model on a triangular lattice in which the only interaction of pseudospins consist in the interdiction to have intersections of domain walls, where domain walls are by definition the lines separating unequal pseudospins which are next-to-nearest neighbors of each other on the original triangular lattice (in other terms, nearest neighbors on the same sublattice). Remarkably enough, an Ising model with such an interaction has been introduced by Fendley, Moore  and Xu \cite{Fendley2007} as a simple example of a two-dimensional model with a local geometric constraint without any reference to some physical system.

For us, the main practical advantage of using the pseudospin representation lies in its convenience for constructing efficient algorithms that sample the space of all domain-wall configurations allowed in the considered Heisenberg model, Eq.~\eqref{eqn:Hamiltonian}. In principle, these configurations
can be then used to study correlations of the Heisenberg spins for any value of $\Phi$ from the interval $(0,2\pi/3)$.
For other physical observables, an even more
straightforward correspondence between the two models exists. Consider, for instance, the chirality $\chi_k$ defined as
\begin{equation}
\label{eqn:chiralityperplaquette} \chi_k=\mbox{sign}\{{\bf S}_{{j}_A(k)}
\cdot [{\bf S}_{{j}_B({k})}\times{\bf S}_{{j}_C({k})}]\}= \pm 1\, ,
\end{equation}
where $j_{\alpha}(k)$ is the site of plaquette $k$ belonging to sublattice
$\alpha$ (where $\alpha=A,B,C$). The chirality always changes sign upon
crossing a domain wall, since one of the spins is flipped with respect to
the plane formed by the other two spins. In terms of the pseudospin
representation, the same quantity is given by
\begin{equation}
\label{eqn:chiralityperplaquette2}
\chi_k={\sigma}_{{j}_A(k)} \sigma_{{j}_B({k})} \sigma_{{j}_C({k})},
\end{equation}
because now one of the pseudospins changes sign upon crossing a domain
wall, while the other two are unchanged. This correspondence suggests that the chiral properties do not depend on the angle $\Phi$.

Even for the magnetic properties, for which there is no such a simple
correspondence, it is possible to conclude on the existence of ordering by
analyzing the properties of the pseudospin model. As discussed in
Ref.~\onlinecite{Korshunov2012}, one expects that the presence of
long-range correlations follows from having only a restricted number of
uncontractable domain walls between two distant points on the lattice.
Thus, studying the statistics of loops (closed domain walls) in the
pseudospin model allows one to reach qualitative conclusions
regarding the behavior of spin-spin correlations in the original Heisenberg
model, as done in Sec.~\ref{sec:MTcluster:Cayleydist}.

In terms of the pseudospin representation, each regular state of the
original Heisenberg model corresponds to one of the states in which on
each sublattice all pseudospins are in a perfectly ordered state. It is
clear that in total there exist only eight such states and that the
infinite multitude of regular states, which can be achieved by successive
spin flips on different sublattices, can be mapped onto these eight states.
Thus each of the eight perfectly ordered states of the pseudospin
representation may correspond to different regular states of the original
model. Accordingly, the equivalence between the two models has some restrictions which are discussed in the next section.

\subsection{\label{sec:Heisenbergsector}Boundary conditions and the Heisenberg sector}

\begin{figure}
\centering
\includegraphics[width=\columnwidth]{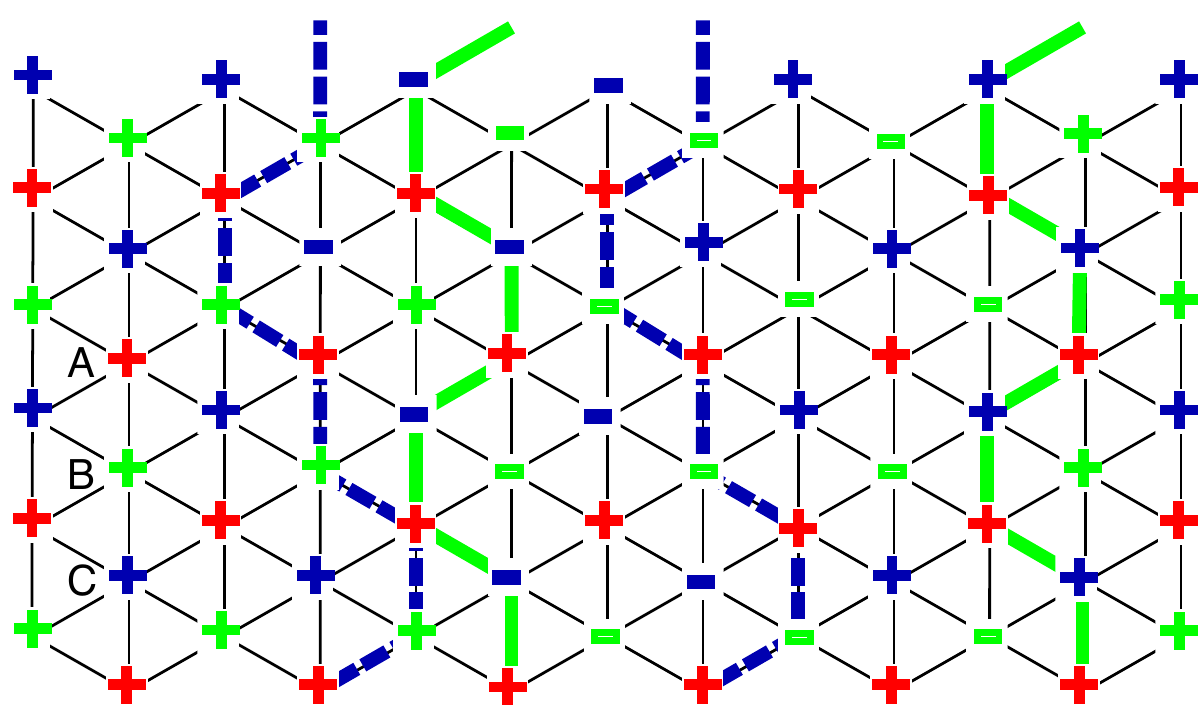}
\caption{\label{fig:windingloops}(Color online) Example of a domain
  wall configuration (thick lines) in the pseudospin model with periodic boundary conditions
  which does not contradict the local rules but which is prohibited in the Heisenberg model. The values of the pseudospins are indicated by the plus and minus signs which are color coded according to the sublattice.}
\end{figure}

The manifolds of all domain-wall configurations allowed by the local rules for the original Heisenberg model and for
the pseudospin model are rigorously identical only when the system has a simple topology,
i.e. lives on a simply connected lattice. However, when one imposes
periodic boundary conditions, an important difference between the two
models appears, as illustrated in Fig.~\ref{fig:windingloops}.

This figure presents a configuration of pseudospins $\sigma_j=\pm 1$
with four domain walls crossing the whole system in the
vertical direction. This configuration is perfectly
compatible with having periodic boundary conditions both in the horizontal
and vertical directions. However, in the case of Heisenberg
spins, such a sequence of domain walls corresponds (when going from left to
right) to successively flipping spins on sublattice $C$, then on sublattice $B$, then again
on sublattice $C$, and finally again on sublattice $B$.
Such a sequence of spin flips results in the rotation of the spins on the
sublattices $C$ and $B$ by an angle
$\Psi(\Phi)=2\arccos\frac{\cos\Phi}{1+\cos\Phi}$ around the direction of
the spins on sublattice $A$.\cite{Korshunov2012} This means that in the
generic case the regular states in the leftmost and rightmost domains of
Fig.~\ref{fig:windingloops} are different. Therefore, in terms of the
Heisenberg model, such a configuration of domain walls is not compatible
with periodic boundary conditions in the horizontal direction.

Thus, in the case of periodic boundary conditions, the manifold of
allowed domain wall configurations in the pseudospin model is larger than
that of the original Heisenberg model. Below we refer to the set of
domain wall configurations of the pseudospin model which are consistent with the periodic
boundary conditions for Heisenberg spins as the {\em Heisenberg sector} of
the pseudospin model. If the system has long-range order
(as expected from the analysis of Ref.~\onlinecite{Korshunov2012}),
the difference between the full partition function of the pseudospin model
and that of its Heisenberg sector has to become negligible in the
thermodynamic limit, since each state not belonging to the Heisenberg
sector involves the presence of at least four domain walls crossing the
whole system, while the presence of long-range order is incompatible with the existence
of infinitely long domain walls.

Quite remarkably, the numerical analysis of the pseudospin model is
automatically restricted to the Heisenberg sector if simulations are
performed with the help of an algorithm based on local updates and are
started from a perfectly ordered configuration. This is precisely the
strategy which is applied in the next section. A more efficient algorithm
involving sampling of all the states of the pseudospin model is introduced
later, in Sec. \ref{sec:MTcluster}.

\section{\label{sec:MTmetro}Metropolis Monte Carlo sampling of ground-state
manifold}
\subsection{Algorithm}\label{sec:algorithm}

We will now describe the most basic Markov-chain Monte Carlo technique
that allows to sample ground-state configurations in the pseudospin
problem. The key observation is that in any ground-state configuration one
can separate all pseudospins into flippable and unflippable. A pseudospin
$\sigma_j$ at site $j$ is said to be flippable if flipping its sign
leads to another ground state with 
nonintersecting
domain walls. For this, there should be no domain wall passing through that
site. On the other hand, pseudospins on sites passed through by domain
walls are unflippable, because flipping any of them would create domain
wall crossings which are prohibited. Naturally, each flippable
(unflippable) pseudospin $\sigma_{j}$ corresponds to a flippable
(unflippable) spin ${\bf S}_{j}$ in terms of the original Heisenberg
model.

Accordingly, the simplest local Monte Carlo move one can imagine consists
in changing the sign of a flippable pseudospin. In detail, our approach
consists in defining a Monte Carlo sweep as a sequential scan through all
the lattice sites with a flippability test of all the spins visited. If a
spin is flippable, the Monte Carlo move of changing its sign is accepted
with $50\%$ probability. An alternative scheme would be to randomly choose
spins and to always accept the move if a chosen spin is flippable.  Note
that both schemes satisfy detailed balance and are ergodic but have the
obvious drawback of any local update -- a slow displacement in the
configuration space resulting in large autocorrelation times.
On the other hand, this simple MC method has the convenient property that all pseudospin configurations generated by using local moves remain in the Heisenberg sector, if we start the Markov process from a regular state.

In the simulations, we construct the triangular lattice in the usual way
by using a square lattice of linear size $L$  (containing $N=L^2$ sites)
with diagonals along one direction. Periodic boundary conditions are
applied with respect to the base square lattice. In such a setup, in order
to keep the three-sublattice structure, $L$ has to be a multiple of $3$. We
have verified the correctness of our algorithm by comparing its results
with those of an exact enumeration approach in which we systematically generated all possible ground states in small systems up to the one containg $N=6^2$ spins (which in the case of periodic boundary conditions has
exactly $2\,176\,672$ degenerate ground states).

\subsection{Magnetization and Chirality}
\label{sec:magnchirdef}

In order to study whether a typical state belonging to the ground
state manifold is ordered, we turn to an estimation of physical
observables (expressed in terms of the pseudospins) using the MC sampling
procedure. The basic magnetic quantities studied below are the sublattice
magnetizations
\begin{equation}
m_{\alpha} = \frac{3}{N} \sum_{i\in {\cal T}_\alpha} \sigma_i\,,
\end{equation}
where $\alpha\in \{A,B,C\}$ denotes a particular triangular sublattice and
${\cal T}_\alpha$ is the set of all sites belonging to this sublattice.  In any
state with perfectly ordered sublattices, these magnetizations are equal
to $\pm 1$ resulting in an 8-fold degeneracy in accordance with the number of possible combinations of signs. In order to verify if the same order parameter
survives in a typical ground state, we also analyze the size dependence of
the mean of the absolute values of the three sublattice magnetizations,
\begin{equation}
 m_s = \frac{1}{3} \sum_\alpha | m_{\alpha} |\,,
\end{equation}
as well as the analogous mean of
the squared sublattice magnetizations. 
In a state in which at least one sublattice has long-range order, both $\langle m_s \rangle$ and $\langle m_\alpha^2 \rangle$ will be nonvanishing and
$\langle m_\alpha^2 \rangle$ can be expected to converge to the same limit
as the spin-spin correlation function $\langle \sigma_i \sigma_j \rangle$,
where $i$ and $j$ are sites on the same sublattice maximally distant from
each other. Here and below angular brackets denote
the standard ensemble average, which is calculated by Monte Carlo sampling.

We also consider the chirality $\chi$ defined
by
\begin{equation}
\label{eqn:chirality}
\chi = \frac{1}{N} \sum_{k\in \triangle} \chi_k\,,
\end{equation}
with $\chi_k$ the chirality on plaquette $k$ as defined in
Eqs.~\eqref{eqn:chiralityperplaquette} and
\eqref{eqn:chiralityperplaquette2}. The sum in Eq.~(\ref{eqn:chirality})
runs over all $N=L^2$ \emph{up} triangular plaquettes of the lattice. \footnote{The sum over \emph{down} plaquettes has the same value.}

\subsection{Simulation results for magnetization and chirality}
\label{sec:magnchirlocal}

We have calculated the above quantities for various linear sizes between $L=12$ and $L=192$ (as explained in Sec.
\ref{sec:algorithm}, $L$ has to be a multiple of $3$). In each case we took
$10^5$ independent measurements. Independence of the measurements was
achieved by calculating observables only for every $L^2$ generated
configurations, reflecting the increase of autocorrelation time $\tau \sim
L^2$ found for the sampling scheme (from time series of the magnetization
and the chirality). The overall time complexity of the local algorithm is
thus of the order of $O(L^4)$.

\begin{figure}
\includegraphics[width=0.9\columnwidth]{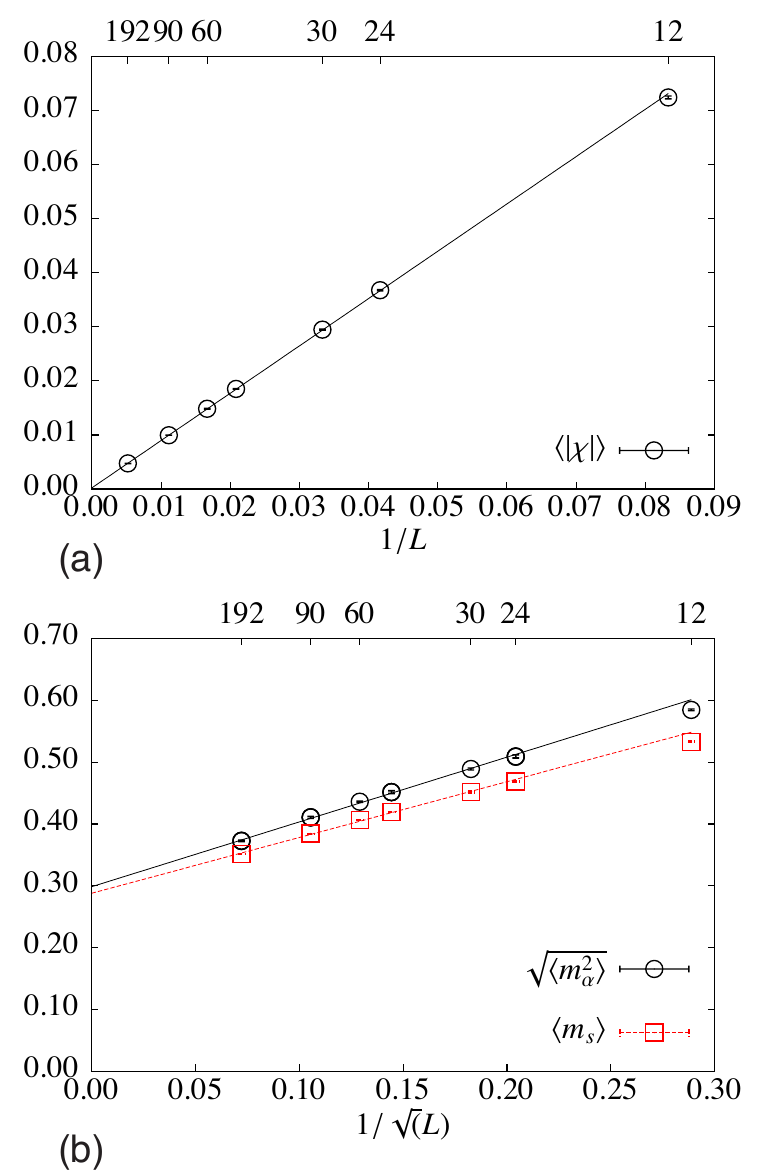}
\caption{\label{fig:MetroMC}(Color online) (a) The chirality $\langle
|\chi| \rangle$ tends to zero like $1/L$.  (b) Size dependence of the
  average sublattice magnetization $\langle m_s \rangle$ and of
  $\sqrt{\langle m_\alpha^2 \rangle}$. In the thermodynamic limit, both
  quantities extrapolate to a same finite value with corrections that
  vanish approximately like $1/\sqrt{L}$. Here, as well as in the subsequent plots, error bars are shown but are smaller than the symbol size. }
\end{figure}

Since in any perfectly ordered state (without domain walls) the normalized
chirality defined by Eq.~\eqref{eqn:chirality} is equal to plus/minus one,
a study of $\langle |\chi|\rangle$ should be one of the most direct and
simple ways to check if a typical ground state is ordered as we expect.
However, our MC results for the chirality, shown in
Fig.~\ref{fig:MetroMC}(a), do not confirm this expectation since they do not show any tendency for the convergence to a finite value. On the contrary, these data rather accurately follow the straight line corresponding to a $\langle |\chi| \rangle\propto 1/L$ decay of the chirality with the system size.
  Note that this dependence should
  not be taken as a sign of algebraic correlations of $\chi_k$.
  Indeed, the same dependence would be obtained if the chiralities of
  different plaquettes were simply uncorrelated from each
  other. Thus, the observed size dependence of $\langle |\chi|\rangle$ suggests that the ground states are
  disordered in terms of chirality, or, at most, have a very small
  chirality whose resolution is beyond the limits of the method used
  in this section.

In evident contrast to this finding, the probe for ordering based on the
sublattice magnetization is clearly in favor of an ordered situation. A
finite-size analysis of both $\langle m_s \rangle$ and $\sqrt{\langle
m_\alpha^2 \rangle}$, shown in Fig.~\ref{fig:MetroMC}(b), indicates the
presence of a long-range ordering with both quantities converging to the
same finite value of the spontaneous magnetization $\langle m_s \rangle \approx 0.29$ in support of the conclusions of Ref.~\onlinecite{Korshunov2012}.
This extrapolation is based on assuming that the finite-size corrections
decay like $L^{-1/2}$, as clearly suggested by the data
presented in Fig.~\ref{fig:MetroMC}(b). The value of the magnetization is
naturally suppressed compared to the totally ordered (regular) states.
Still, under the assumption that in the ordered state all three
sublattices remain equivalent, the simplest estimate assuming $\langle
|\chi| \rangle \sim \langle m_s \rangle^3$ would suggest for the chirality a
detectable value $\langle |\chi| \rangle \sim 0.024$ which is clearly not
observed.

\begin{figure}
\includegraphics[width=0.9\columnwidth]{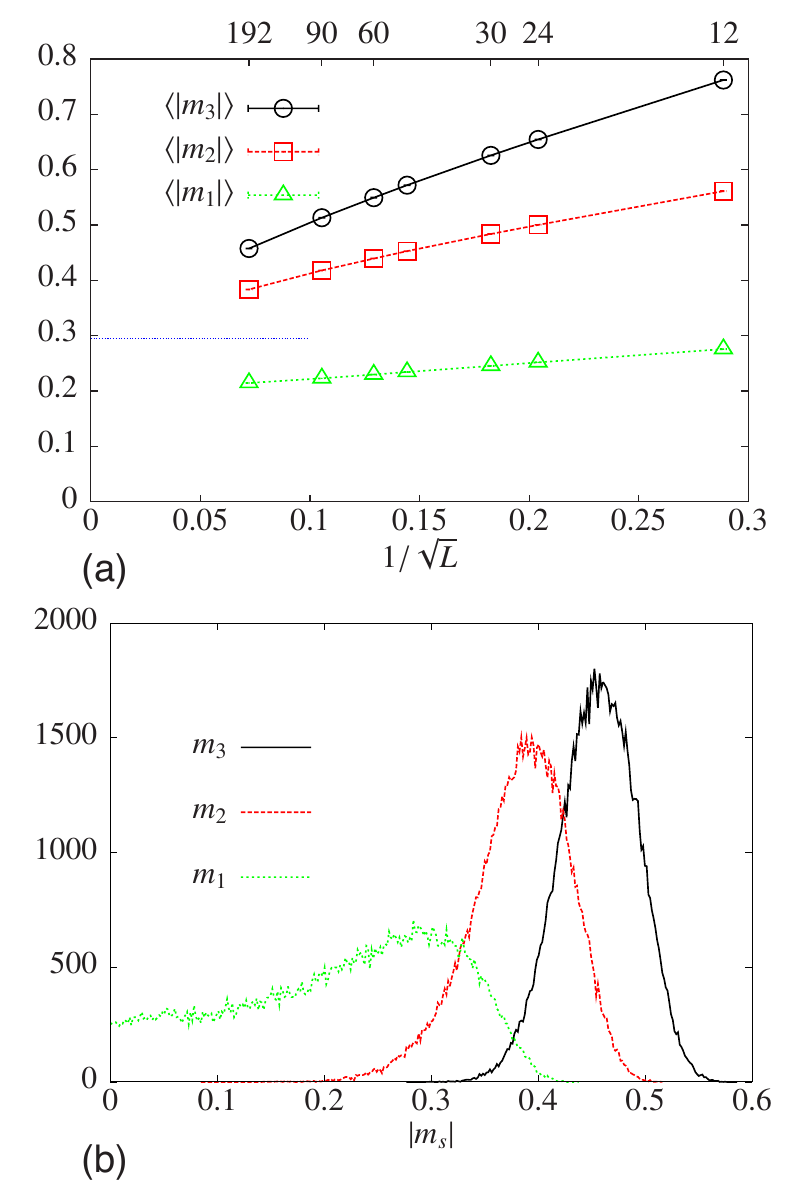}
\caption{\label{fig:MetroMCHist}(Color online) (a) Size dependence
  of the three sorted sublattice magnetizations $\langle |m_{1}|
  \rangle$, $\langle | m_{2} | \rangle$, and $\langle | m_{3}
  | \rangle$.  While the behavior of $\langle |m_{3}| \rangle$,
  $\langle | m_{2} | \rangle$ is consistent with an extrapolation
  towards the average sublattice magnetization (shown with a dashed horizontal line),
  the behavior of
  $\langle |m_{1}| \rangle$ is puzzling. Lines are included as a guide to the eye. (b) Distribution
  functions of the sorted sublattice magnetizations encountered in
  ground-state configurations for $L=192$, indicating the frequent
  presence of partially disordered configurations.}
\end{figure}

Motivated by this lack of consistency, we decided to have a more attentive
look at the distribution and the size dependence of individual sublattice
magnetizations to check if in a typical ground state the three sublattices
are equivalent, that is have (up to a sign) the same magnetization in the
thermodynamic limit. Any other situation would mean the presence of an
additional discrete degeneracy related to the breaking of the symmetry
with respect to the permutation of sublattices. To this end, we record for
each ground-state configuration the triad of sublattice magnetizations
$\{m_{A},m_{B},m_{C}\}$ and sort them into an ordered triad $\{ m_{1},
m_{2}, m_{3} \}$ with $|m_{1}| \leq | m_{2} | \leq | m_{3}|$. Figure
\ref{fig:MetroMCHist}(a) shows MC results for the sorted magnetizations,
which make apparent that the sublattice order is to a large degree
heterogeneous: Typically two of the three sublattices show ordering
correlations while the third has a much smaller magnetization together
with a finite-size scaling trend which seems incompatible with an
ordered state in which all three sublattices order (up to the sign)
equally. This is also evident in Fig.~\ref{fig:MetroMCHist}(b) showing
typical distribution functions of the sorted magnetizations $|m_{1}|$,
$|m_{2}|$ and $|m_{3}|$ for $L=192$, indicating that $m_{1}$ has a very
prominent support (large weight) in the region of disordered
configurations.

These data are clearly insufficient to understand whether  (i) a typical ground state
is partially ordered with one sublattice becoming completely disordered in the thermodynamic limit (leading to a state trivially consistent with a vanishing chirality), or
(ii) when $L$ increases, the minimal magnetization $m_{1}$ will eventually reverse its trend and converge towards the same limit as $m_{2}$ and $m_{3}$, or (iii) some other (intermediate) scenario is realized.
In any case, the question whether a finite chirality
is ultimately stabilized seems primarily linked to the fate of the
smallest sublattice magnetization $m_{1}$ in the thermodynamic limit and
the ambiguity in the current data calls for simulations on much larger
systems in order to resolve this question. The remainder of this paper is
focused on the construction of an efficient algorithm allowing to
undertake such simulations as well as on its implementation for a
numerical resolution of the puzzle just described.

\section{\label{sec:MTcluster}Cluster Monte Carlo Algorithm}

\begin{figure}
\includegraphics[width=\columnwidth]{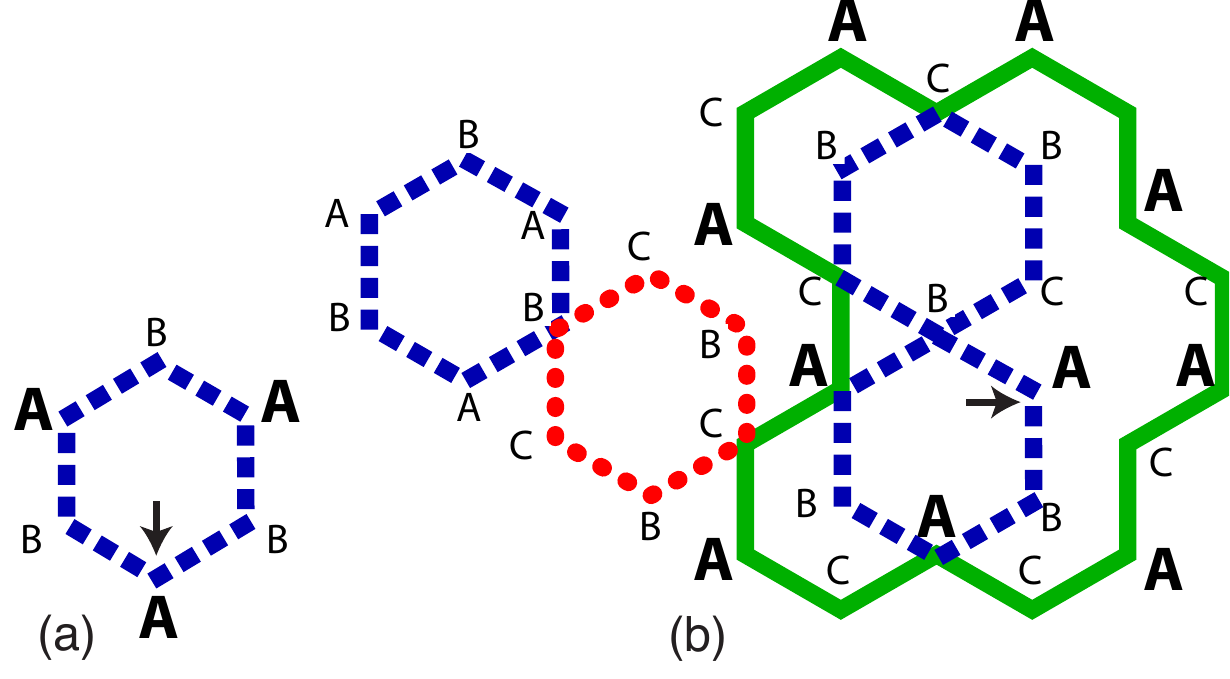}
\caption{\label{fig:cluster}(Color online) Visualization of flippable
clusters of pseudospins. 
(a) A simple cluster of ${\bf A}$ spins (shown in bold) within an $AB$ loop that does not touch any other loops. (b) A more complicated cluster of ${\bf A}$ spins (shown in bold) in a situation where multiple domain walls touch each other. Note that not all $A$ sites of the domain wall network belong to the flippable cluster. In both panels, the site from which the construction of the flippable cluster is started is indicated by an arrow.}
\end{figure}

In the local MC procedure used in Sec.~\ref{sec:MTmetro}, the updates were
restricted to individual pseudospins which can be flipped without creating
domain walls crossings. Such \emph{flippable} spins are easily
recognizable because they sit on the sites which are not passed through by
any domain wall. Locally flipping an \emph{unflippable} spin (sitting on a
site belonging to some domain wall) would result in a violation of the
constraint that domain walls cannot cross. It is, however, possible to
flip an unflippable pseudospin by flipping a whole cluster of unflippable
pseudospins from the same sublattice, and it will be natural to call such
a cluster a {\em flippable cluster}. A flippable pseudospin can be
considered as the simplest example of a flippable cluster. The first
non-trivial example of a flippable cluster (e.g.~on sublattice ${\cal
T}_A$) consists of all pseudospins on ${\cal T}_A$ which are part of, say,
an $AB$ loop not touching another domain wall, see
Fig.~\ref{fig:cluster}(a). To avoid the intersections of domain walls, all
pseudospins on ${\cal T}_A$ visited by the same wall have to be equal to
each other. Naturally, after being flipped they remain equal to each
other, which further ensures the  absence of wall-crossings.

Flippable clusters have a more complicated structure when domain walls touch
each other. Consider for instance a particular $AB$ loop touching an $AC$
loop (or a few of them). To ensure the absence of wall-crossings one has to
flip not only all ${\cal T}_A$ pseudospins belonging to a chosen $AB$
loop, but also all ${\cal T}_A$ pseudospins belonging to the $AC$ loops
touching it, and also all ${\cal T}_A$ pseudospins belonging to all $AB$
loops touching these $AC$ loops and so on. In other words, one has to flip
${\cal T}_A$ pseudospins on the whole connected cluster formed by $AB$ and
$AC$ loops touching each other, as indicated in Fig.~\ref{fig:cluster}(b).
Note that in terms of the original Heisenberg model all pseudospins
belonging to the same flippable cluster correspond to equal spins ${\bf
S}_{i}$.

Using this approach, a valid global update can be constructed applying a
single cluster technique in the spirit of the Wolff algorithm,
\cite{Wolff89, SwendsenWang97}
but with completely deterministic cluster construction:

\begin{itemize}
\item pick a random site $i$ (and determine its sublattice index $\alpha$,
say $A$); \item check if spin $\sigma_i$ is flippable and flip it if it is;
\item if $\sigma_i$ is unflippable, construct the connected cluster of
domain walls formed by $AC$ and $AB$ loops and flip all the spins on
sublattice $A$ belonging to that cluster.
\end{itemize}

The construction of connected clusters can be done purely recursively in a
depth-first approach which is usually the most convenient choice to code.
 However, this technique requires large stack memory
resources for big clusters which can be a severe limitation. In view of
the large system sizes that are required in this study, we have opted for
a hybrid cluster identification approach in the spirit of
Ref.~\onlinecite{Herrero2004JPhA} in which we follow domain walls
\emph{iteratively} until a touching point with another domain wall is
found in which case we continue cluster construction recursively. Such an
approach is expected to be less demanding with respect to stack memory. In our
implementation, a typical Monte Carlo sweep then consists of one pass
through the lattice where we check for flippable spins and apply the local
update.  This sweep is followed by $N_{\rm{sc}}$ single-cluster updates
where $N_{\rm{sc}}$ is an integer number such that on average say
$N/3$ sites are touched during the cluster construction (in an
accumulated way). This number $N_{\rm{sc}}$ is determined
self-consistently during the equilibration phase of the algorithm.  With
this setup, the nonlocal update turns out to be highly efficient with no
sign of critical slowing down at all: the dynamic exponent $z$ [measured
from the size dependence of the autocorrelation time $\tau(L)\sim L^z$ for
the total magnetization $m=\sum_\alpha \sigma_\alpha$] cannot be
distinguished from zero. Essentially this means that the algorithm
generates independent configurations with time complexity $O(L^2)$
compared to $O(L^4)$ for the local update.

\section{\label{sec:MTcluster:results}Simulation Results}
Using the cluster algorithm, we have investigated much larger systems than
before going up to $L=11340$ which is roughly $50$ times larger than
could be done with the local update.  In detail, we studied systems of
linear sizes $L=384, 576, 1152, 2304, 3456, 4608, 5760, 11340$ next to
those already considered in the previous section (the total number of
spins $N$ being $L^2$, there were $128\,595\,600$ spins in the largest cluster). Typically we collected of the order of $10^6$
independent measurements (for the largest system sizes $10^5$) in each
case.

\subsection{\label{sec:MTcluster:winding}Winding loops and the Heisenberg sector}

As already mentioned in Sec.~\ref{sec:Heisenbergsector}, a special
property of the cluster update (in contrast to the local MC procedure) is
that in a system with periodic boundary conditions
it can generate pseudospin configurations which do not correspond to any ground state of
the original Heisenberg model with the same boundary conditions.\footnote{In the case of free boundary conditions this discrepancy does not
 exist.} All such
states are characterized by the presence of winding domain walls, that is,
of incontractible loops which wind around the torus. However,
not all ground states containing some sequence of winding domain walls
correspond to the violation of the periodic boundary condition in the
Heisenberg model. Some sequences of winding domain walls (which below are
called {\em reducible}) can be obtained as well in the framework of local
update and the states containing them represent true ground states of
the Heisenberg model.

In order to check if a given pseudospin configuration belongs to the
Heisenberg sector or not (in the latter case it contains an {\em
irreducible} sequence of winding domain walls) the following simple
procedure can be used. Let us denote a sequence of domain walls crossed by
a closed path going around one of the two directions on the torus by a
word $\mathsf{w}$, for example $\mathsf{w}=\text{CABBAC}$, in which each
letter stands for the type of domain wall crossed by the path. Namely,
letter A stands for crossing a loop consisting of segments along $BC$
links, B for segments along $CA$ links, and C for $AB$ type wall. A simple
check whether the configuration is compatible with the Heisenberg sector
consists in testing whether the word $\mathsf{w}$ obtained in such a way
can be reduced to an empty word by repeatedly applying the rule that two
successive letters in the word annihilate themselves if they are
identical. The possibility to get rid of two identical letters when they are adjacent originates from the possibility to annihilate a pair
of neighboring domain walls of the same type by making local single-spin
flips on the corresponding sublattice between them. For instance, the
word $\mathsf{w}=\text{CABBAC}$ can be reduced to the empty word (since
$\text{CABBAC}\rightarrow\text{CAAC}\rightarrow \text{CC}$) whereas the
word CBCB, obtained from Fig.~\ref{fig:windingloops} when going along a
closed path in the horizontal direction, cannot and is thus said to be
irreducible. A configuration belongs to the Heisenberg sector if for both
directions on the torus the procedure described above produces empty
words.

Below, the fraction of configurations outside the Heisenberg sector is
denoted by $p_{\rm nH}$ and the fraction of configurations with reducible
sequences of winding domain walls by $p_{\rm w}$.  An analysis of both
quantities is of interest as in general all sequences of winding domain
walls generate competing strips of domains spanning the whole system.
Hence, even the Heisenberg sector can suffer from considerable finite-size
effects due to the presence of reducible sequences of winding domain
walls. In particular, since domain walls cannot cross each other, the
winding domain walls have to choose one of the two preferred directions
imposed by periodic boundary conditions and therefore their presence makes
a configuration essentially anisotropic.

\begin{figure}
\includegraphics[width=0.9\columnwidth]{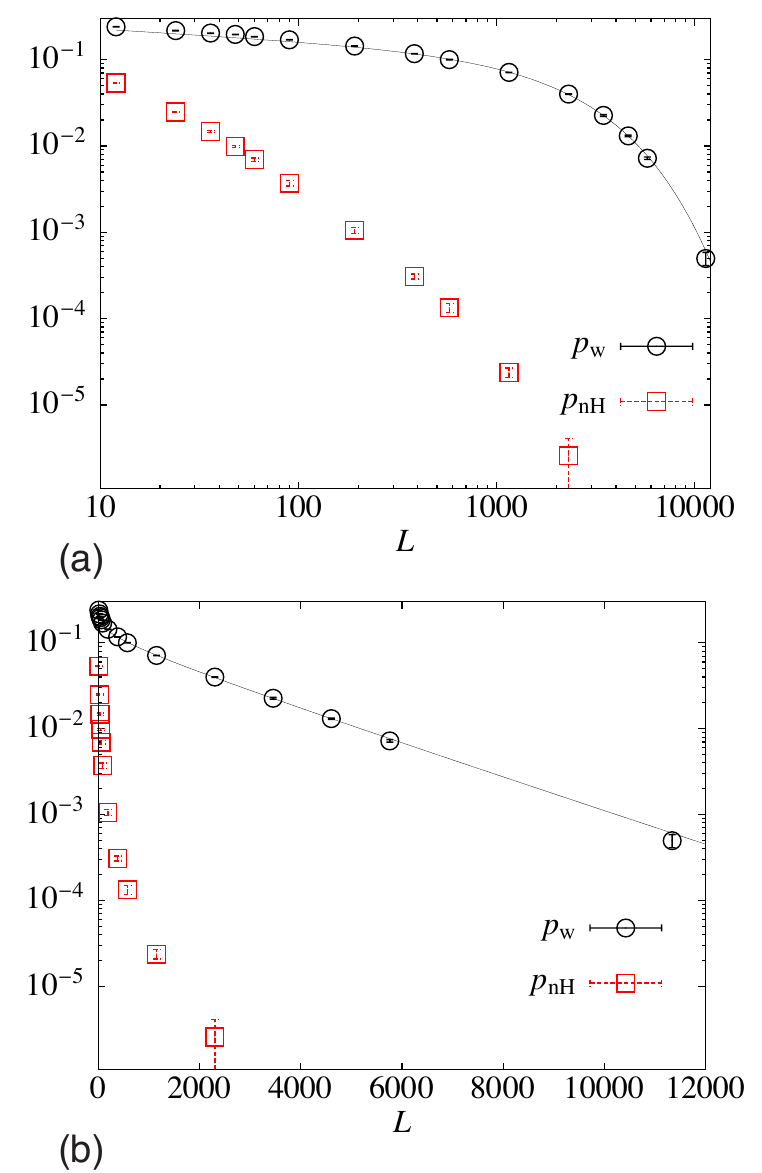}
\caption{\label{fig:nontrivialsector}(Color online) The proportion of domain-wall
  configurations that are inconsistent with the Heisenberg sector
  ($p_{\rm{nH}}$) quickly drops to zero as a function of lattice size as
  shown in a double logarithmic plot (a) and in a semilogarithmic plot (b).
  For $L>2000$ no such configurations were observed within the
  simulation time.  On the other hand, a much larger fraction $p_{\rm w}$
  of configurations belonging to the Heisenberg sector but incorporating
  winding loops is observed to decay much more slowly.
  For sizes $L\gsim 1000$ the behavior of $p_{\rm w}$ is consistent
  with an exponential decay as evidenced in the semilogarithmic plot.}
\end{figure}

First, for the configurations outside the Heisenberg sector, the
simulations undoubtedly demonstrate that $p_{\rm nH}$ very quickly drops
to zero, see Fig.~\ref{fig:nontrivialsector}. Therefore, these
configurations have no effect whatsoever on physical quantities in the
thermodynamic limit and the results obtained in the framework of cluster
update are applicable also for the description of the original Heisenberg
model.
Even in small systems with $L=12$, where we find a considerable fraction
(about $\approx 0.1$) of such non-Heisenberg configurations, the
systematic error made on estimating physical quantities without filtering
out the non-Heisenberg configurations is only of the order of $0.4\%$ for
the chirality and $1.8\%$ for the sublattice magnetization. Naturally,
restricting the analysis to the Heisenberg sector, we recover the same
results as with the local update.

A finite-size analysis of $p_{\rm w}$, also shown in
Fig.~\ref{fig:nontrivialsector}, highlights the fact that the fraction of configurations with reducible sequences of winding domain walls remains considerable up to very large system sizes, with a rather slow decrease described
initially by an algebraic-like behavior with a small effective exponent
$b\approx 0.2$. For lengths greater than $\sim 1000$, the decrease,
however, is consistent with an exponential decay $\sim \exp(-L/\lambda)$.
We determine this length scale to be given by $\lambda\approx 2000$. The
exponential decay of $p_{\rm w}$ suggests that all domain walls have a
finite free energy per unit length which is compatible with having a true
long-range order on all three sublattices.

\subsection{Solving the ordering puzzle}
We are now in a better position to readdress the puzzling situation
described in  Sec.~\ref{sec:magnchirlocal}, that is, to study whether the
non-coplanar ground states have true long-range order (in magnetization
and in chirality), are only partially ordered, or if some entirely
different scenario is realized. To learn more on the nature of sublattice
ordering, we start by extending the analysis of the size dependence of the
ordered triad of magnetizations $\{ m_{1}, m_{2}, m_{3} \}$ (presented for
$L\le 192$ in Fig.~\ref{fig:MetroMCHist}) to larger systems.
\begin{figure}
\includegraphics[width=0.9\columnwidth]{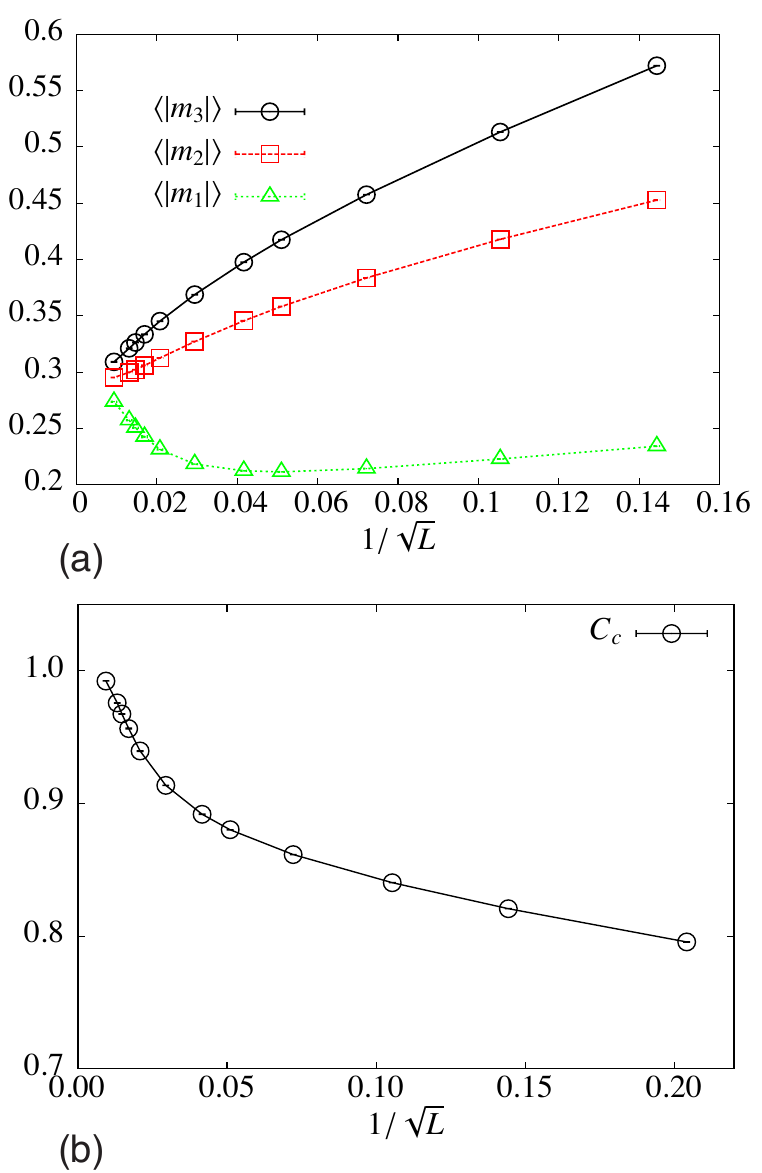}
\caption{\label{fig:cubicity}(Color online) (a) Plot of the sorted
  sublattice magnetizations  for $L$ up to 11340 demonstrating that the
  three sublattices become equivalent in the thermodynamic limit.
  (b) $C_c$ defined by Eq.~\eqref{eqn:cubicity} provides a quantitative
  measure allowing one to characterize the nature of ordering. Its
  convergence to $1$ (the uppr boundary for $C_c$) indicates the tendency of the sublattice magnetizations
  to become equal to each other. The scale $1/\sqrt{L}$ is somewhat
  arbitrary and has been motivated by the size dependence of the magnetization
  at small scales.}
\end{figure}
The new data, shown in Fig.~\ref{fig:cubicity},
immediately confirm the need for simulations on large $L$: The
behavior of the smallest magnetization $m_{1}$, one of the main puzzles of the local update results, turns out to be non-monotonous. This quantity passes through a minimum (situated roughly at $L\sim 500$) and after that it ultimately tends towards a common limit with $m_2$ and $m_3$. This finding definitely speaks in favor of a situation in which all three sublattices are ordered to an equal degree.

To investigate this question more quantitatively, we study the ratio,
\begin{equation}
\label{eqn:cubicity}
C_c= \frac{3(m_1^2 m_2^2 + m_1^2 m_3^2 + m_3^2 m_2^2)}{(m_1^2 + m_2^2 + m_3^2)^2}\,,
\end{equation}
of the two quartic invariants  normalized in such a way that $C_c$ always
belongs to the interval $[0,1]$. In the thermodynamic limit, $L\to\infty$,
$C_c$ has to tend to $1$ for an ordered state in which all sublattices are, up to the sign, equally ordered, to $3/4$ for a partially ordered
state in which two sublattices are ordered equally and the third one
is disordered, and to $0$ when only one sublattice is ordered.
The size dependence of $C_c$ demonstrates equal strength
sublattice ordering but again clearly points to the difficulty of the
problem: While the data for $L<500$ could be interpreted as tending to
a value around $0.9$, which would imply that all three sublattices are ordered, but have different magnetizations, a relatively abrupt change of regime takes place at a particular length scale with convergence to the state in which all sublattices are equally ordered at larger $L$.

\begin{figure}
\includegraphics[width=0.9\columnwidth]{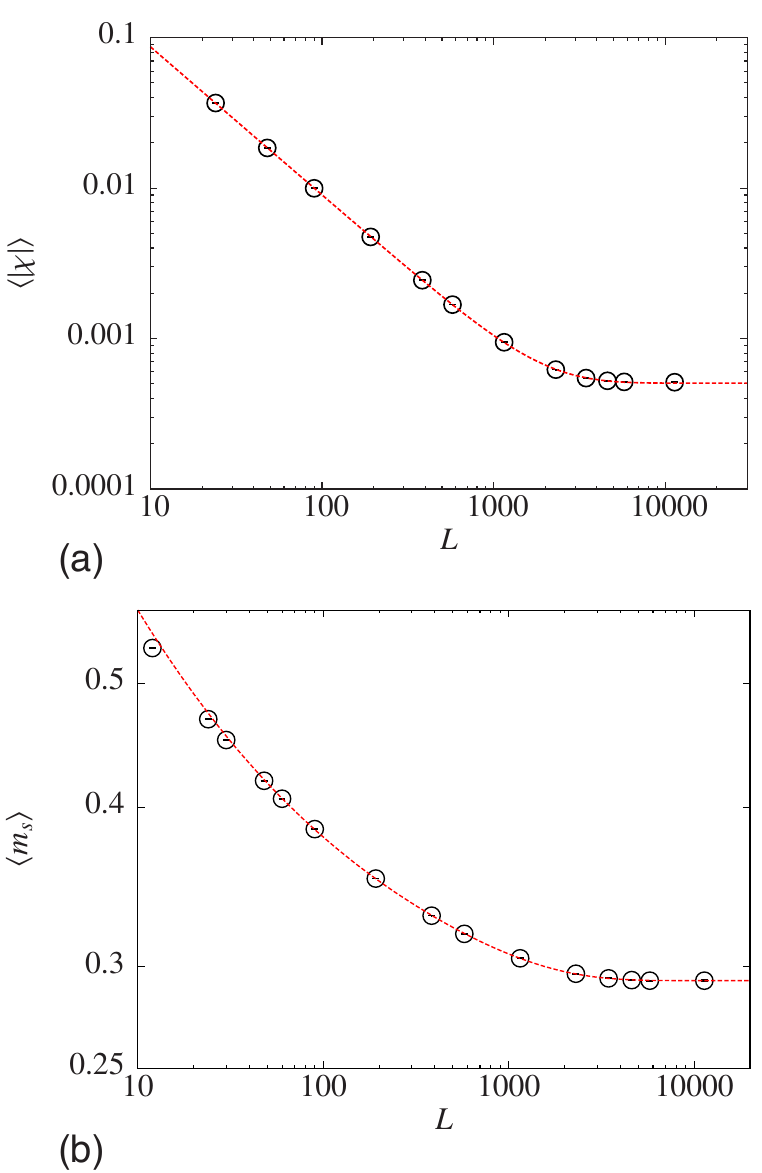}
\caption{\label{fig:ClusterResults1}(Color online) Plot of (a) the
  chirality and (b) the sublattice magnetization in a double logarithmic scale
  covering the complete range of lattice sizes simulated. Both
  quantities show a convergence to their infinite-size values for the
  largest systems studied. Dashed lines are results of fits to Eq.~\eqref{eqn:ffsansatz}.}
\end{figure}

Having thus constrained the possible ordering scenario, the next
non-trivial question is whether the ground-states are characterized in the
thermodynamic limit by a nonvanishing chirality $\chi$. Naturally, we used
the large-scale data also to study the behavior of the chirality, as well as
of the averaged sublattice magnetization to reach a precise
quantitative characterization for the degree of ordering. The numerical
data for $\langle |\chi| \rangle$, shown in a double logarithmic plot in
Fig.~\ref{fig:ClusterResults1}, provide the second key result of this
section. They unambiguously demonstrate the chiral nature of the
non-coplanar ground states, whereas the finite-size behavior again shows
the non-trivial features characteristic for this study: over three decades
in $L$, the decay of chirality very accurately follows an algebraic law
$\langle|\chi|\rangle\sim L^{-1}$ [consistent with the observation in
Fig.~\ref{fig:MetroMC}(b)] before overcoming a length scale upon which a
very small but finite expectation value is reached. Essentially, this
situation is reminiscent of very weak ordering close to a critical point.

Before analyzing this situation in more detail, we note that a similar
observation holds for the sublattice magnetization (concentrating here on
the quantity $\langle m_s \rangle$) which is saturating above a certain
length scale and that the system sizes studied are now sufficiently large
to represent the thermodynamic limit:
The estimates for the sublattice magnetization for $L=5760$ and $L=11340$
agree within the error bar, and take the value $\langle m_s \rangle \approx 0.2926$.
This reinforces the observation that we have reached a length scale in our
simulations where the effective finite-size corrections are no longer
algebraic, but rather have a conventional exponential form $\sim
\exp(-L/\lambda)$, where $\lambda$ is a length scale proportional to the
correlation length.

In order to extrapolate the data as well as to characterize and capture
the corrections over a large window of system sizes, a reasonable ansatz
for the finite-size dependence of some observable $A$ is
\begin{equation}
\label{eqn:ffsansatz} \langle A \rangle =\langle A \rangle_\infty + a
L^{-b} \exp(-L/\lambda)\,,
\end{equation}
where $b$ is an effective exponent and $a$ some constant coefficient. Note
that this form has been chosen just as an effective means of extrapolating
and describing our results, but we are not aware of a particular theory
predicting the finite-size dependences of  quantities of interest for the
present model. For the conventional Ising model on a square lattice, we
have verified numerically that the extrapolation of the magnetization
using Eq.~\eqref{eqn:ffsansatz} close to criticality gives excellent
agreement with exact results.

\begin{table}[t]
\centering \caption{\label{tab:2DIsing}Results of extrapolations of the
 magnetization and chirality based on Eq.~\eqref{eqn:ffsansatz} for
 fitting windows with different $L_{\rm{min}}$. In both parts of the table the two last lines show 
the parameters of a purely exponential fit with $b=0$.}
\begin{tabularx}{\columnwidth}{ l | X | X | c | c | c | c}
\hline\hline
$L_{\rm{min}}$ & $\langle m_s \rangle_\infty$ & $\lambda$ & $a$ & $b$  &   $\chi^2/\rm{d.o.f}$ \\
 \hline
 $100$ & $0.29252(9)$ & $1413(40)$ & $0.96(3)$ & $0.50(1)$ & $1.2$ \\
 $200$ & $0.29252(10)$ & $1412(50)$ & $0.95(5)$ & $0.50(2)$ &  $0.7$ \\
 $500$ & $0.29253(10)$ & $1362(116)$ & $0.81(20)$ & $0.47(6)$ &  $0.5$ \\
 \hline
 $1000$ & $0.29259(8)$ & $986(21)$ & $0.039(1)$ & - &  $0.8$ \\
 $2000$ & $0.29252(10)$ & $1088(50)$ & $0.031(3)$ & - &  $0.4$ \\
 \hline\hline
$L_{\rm{min}}$ & $\langle |\chi|\rangle_\infty$ & $\lambda$ & $a$ & $b$  &   $\chi^2/\rm{d.o.f}$ \\
\hline
$200$ & $0.000511(3)$ & $1546(50)$ & $0.66(5)$ & $0.94(2)$ &  $5.3$ \\
$500$ & $0.000512(1)$ & $1471(60)$ & $0.5$ & $0.87(2)$ &  $2.5$ \\
\hline
$1000$ & $0.000515(3)$ & $1000(60)$ & $0.0017(1)$ & - &  $10.0$ \\
$2000$ & $0.000513(1)$ & $959(20)$ & $0.00140(5)$ & - &  $0.57$ \\
\hline\hline
\end{tabularx}
\end{table}

Table \ref{tab:2DIsing} shows some results obtained from fits for various
fitting windows $L\in [L_{\rm{min}},L_{\max}]$ with fixed
\makebox{$L_{\max}=11340$}. Error bars were obtained using a bootstrap
approach where fits were repeated several times varying the data within
their statistical uncertainty.  In the case of the sublattice magnetization,
very good fits -- in terms of the standard chi-square
test\cite{press2007numerical} for the variance of residuals per degree of
freedom ($\chi^2/\rm{d.o.f}\sim 1$) -- are obtained describing the data
over a wide range of system sizes. For $\langle m_s\rangle$, the parameter
$\lambda$ turns out to be $\lambda\approx 1500$, quantifying thus the
order of magnitude of system sizes to be overcome in the simulations, and the effective
exponent $b\approx 0.55$ agrees well with the approximate finite-size
behavior $~L^{1/2}$ observed in the simulations using purely local Monte
Carlo moves. In addition to fits to Eq.~\eqref{eqn:ffsansatz}, we have
performed fits using purely exponential corrections ($b=0$) which work
very well for the largest system sizes $L>1000$ but do not capture
finite-size corrections for smaller sizes, see Tab.~\ref{tab:2DIsing}.
Results from purely exponential fits give consistent values for the
sublattice magnetization and a slightly smaller $\lambda\approx 1000$.

Similarly, the finite-size dependence of the chirality is well described
by $\lambda\approx 1500$ and $b\approx 0.9$ [Eq.~\eqref{eqn:ffsansatz} for
$L>200$] and $\lambda\approx 1000$ (purely exponential fit for $L>2000$)
but the quality of the fits is generally less good than for the
magnetization indicating that additional finite-size effects are likely to
be present. However, the finite-size extrapolated value of the chirality
is rather stable and consistent between all the fits.

In summary, the large-scale simulations in the present section reveal not
only the nature of magnetic order (with equal magnetization of all three
sublattices) in a typical ground state but also establish the chiral
nature of such a state, essentially resolving the puzzling situation
described in Sec.~\ref{sec:MTmetro}. Our final estimates for the
infinite-size local order parameters characterizing the order in the
pseudospin model are
\begin{align}
\label{eqn:result:ms}
\langle  m_s  \rangle_\infty &= 0.2925(2)\\
\langle | \chi | \rangle_\infty &= 0.000513(5)\,
\end{align}
where the error bar is a conservative estimate (twice as large as the
largest error bars from the fits). Interestingly, the chirality turns out
to be reduced by two orders of magnitude compared to the naive estimate
$\langle  m_s \rangle_\infty^{3}\approx 0.025$ indicating the presence of
large fluctuations.

\subsection{\label{sec:MTcluster:Cayleydist}Irreducible sequences of domain
walls, distribution on the Cayley tree, and correlation functions}

The cluster update has enabled us to conclude on the nature of ordering in
the pseudospin model by obtaining a precise characterization of the
magnetic order as well as by proving the chiral nature of a typical ground
state. Since the chirality can be also defined in terms of the Heisenberg
spins, the latter conclusion directly applies to the original Heisenberg
model, but the relation between magnetic properties of the two models
is less straightforward. In this Section, we suggest an approach to
characterize quantitatively fluctuations of the Heisenberg model in its
full order-parameter space (without reduction to the pseudospin
representation) and use it to draw conclusions directly on the magnetic
ordering of Heisenberg spins.

Following the presentation of Ref.~\onlinecite{Korshunov2012}, we remind
the reader that the relationship of any two regular spin domains can be
characterized by the shortest sequence of domain walls one has to cross in
order to get from one domain into the other (the irreducible domain wall
sequence), or, in other terms, by sequentially applying the spin flips
associated with the types of domain walls separating them. For two
given plaquettes on the triangular lattice, $k_1$ and $k_2$, this sequence
is easily obtained by connecting these plaquettes by an arbitrary path on
the dual lattice and constructing a word $\mathsf{w}$ from letters A, B, C
in accordance with the types and order of domain walls which are crossed
along this path. Following the discussion of
Sec.~\ref{sec:MTcluster:winding}, this word has then to be simplified
iteratively applying the rule that any neighboring identical letters
annihilate each other until no further simplification is possible. The
irreducible word ${\mathsf{w}}_{k_1; k_2}$ obtained in such a way is
unique (i.e.~it does not depend on the particular path connecting $k_1$
and $k_2$) as soon as the ground-state configuration belongs to the
Heisenberg sector.
The subsequent analysis is thus
restricted to this sector, although in the thermodynamic limit this is of
no importance, as already discussed in Sec.~\ref{sec:MTcluster:winding}.

Because in an irreducible domain wall sequence adjacent walls
have to be of different types the manifold of all irreducible domain wall
sequences 
has the structure of a tree, namely that of an infinite Cayley tree with
coordination number $3$. This means that when considering a particular
ground state, one can characterize the position of any domain in the order
parameter space by a node on this tree. Accordingly, the length of the
irreducible word, $d=|\mathsf{w}_{k_1; k_2}|$, naturally defines the
distance between the two domains in the order parameter space (that is, on
the Cayley tree). Arguments that this distance remains finite regardless
of the \emph{spatial distance} of the domains was one of the key steps
towards making the case for the existence of magnetic ordering in the
non-coplanar ground states in Ref.~\onlinecite{Korshunov2012}.

\begin{figure}[t!!]
\includegraphics[width=0.9\columnwidth]{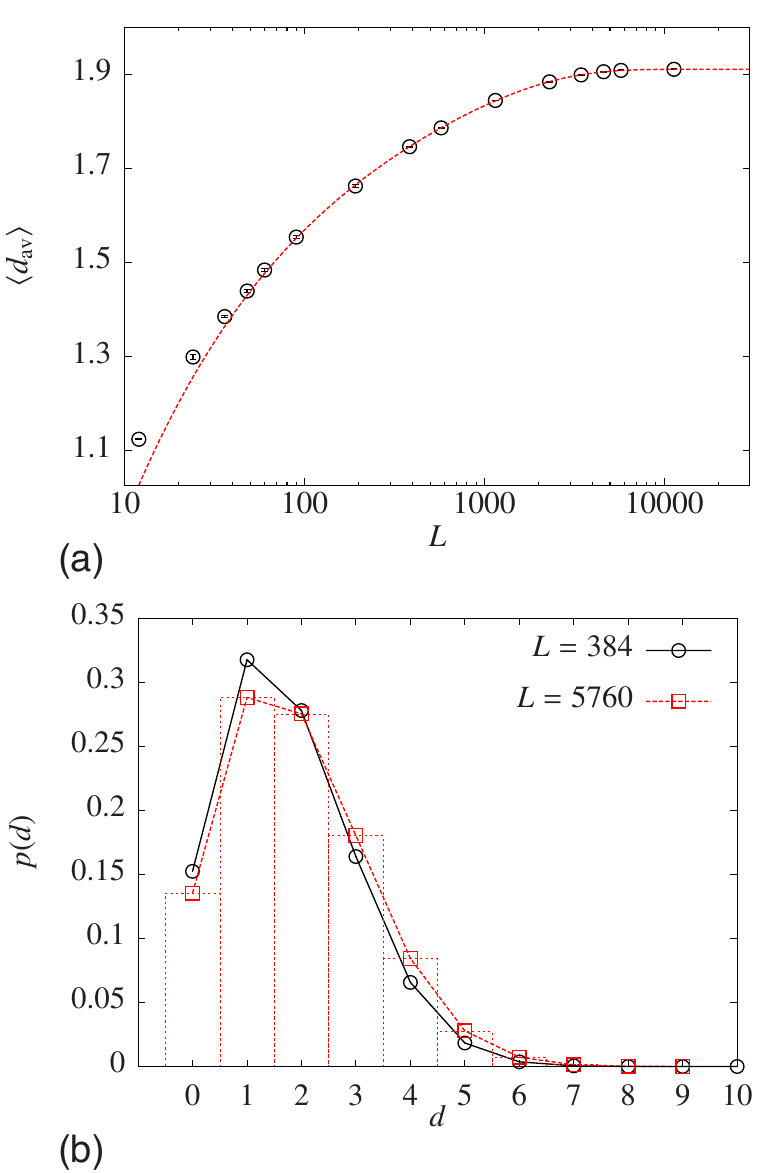}
\caption{\label{fig:Cayleydist}(Color online) (a) Size
  dependence of the average number of irreducible domain walls
  between two plaquettes $k$ and $k^\prime(k)$ situated at the maximal possible
  distance from each other. For very large system sizes a
  convergence is evident. The dashed line is a fit to the ansatz given by
  Eq.~\eqref{eqn:ffsansatz}. (b) Distribution function $p(d)$ of the number
  of irreducible domain walls for two system sizes showing that the
  distance on the Cayley tree $d$ is heavily bounded from above. }
\end{figure}

In order to address this point numerically, we consider plaquettes $k$ and
$k^\prime(k)$ that are maximally separated on the triangular lattice with
periodic boundary conditions [that is are transformed into each other by a
shift by a vector ${\bf{T}}=(L/2, L/2)$] and calculate the average
distance $d_{\rm{av}}$ on the Cayley tree between the domains to which
they belong by using
\begin{equation}
d_{\rm{av}} = \frac{1}{N} \sum_{k} |{\mathsf{w}}_{k;k^\prime(k)}|,
\end{equation}
where the sum runs over all triangular plaquettes with upward orientation. Monte Carlo calculations of
$\langle d_{\rm{av}} \rangle$ show that the average number of irreducible
domain walls is indeed surprisingly small and definitely saturates at a
finite value, see Fig.~\ref{fig:Cayleydist}(a). On average, only
\begin{equation}
\langle d_{\rm{av}} \rangle_\infty = 1.9111(7)\end{equation} irreducible
domain walls separate the domains at $k$ and $k'(k)$ in the thermodynamic
limit, when the distance between $k$ and $k'(k)$ tends to infinity. This
estimate was obtained using finite-size extrapolations according to
Eq.~\eqref{eqn:ffsansatz}, where in particular, for $L$ in the interval
$[192,11340]$ one obtains the following fit parameters:
$\lambda=1900(200)$, $b=0.45(5)$, $\chi^2/\rm{d.o.f}=1.4$.\footnote{In
  the same interval, a fit assuming purely algebraic decay
  of the correction, $g(x)=g_1 +
  g_2 L^{-g_3}$, has a $\chi^2/\rm{d.o.f}=300$ and a fit to a
  polynomial $p(x)=p_1 + p_2/L + p_3/L^2 + p_4/L^3$ has
  $\chi^2/\rm{d.o.f}=50$ highlighting that \eqref{eqn:ffsansatz}
  provides a much better effective description.}
Note that the value of the parameter $\lambda$ is somewhat larger than for the magnetization and chirality. The maximal number of irreducible domain walls ever observed
during our simulations was $10$, see Fig.~\ref{fig:Cayleydist}(b).

Thus, the simulation results presented in Fig. 9 clearly support the
conclusions of Ref.~\onlinecite{Korshunov2012} and confirm that spins
${\bf S}_{k}$ and ${\bf S}_{k^\prime(k)}$ belonging to the same sublattice
indeed remain correlated, irrespective of the angle $\Phi$, even when the
distance between them tends to infinity. This allows one to claim the
existence of a long-range order in the Heisenberg model without making any
reference to the pseudospin model.

Note that, in general, distribution functions such as the one shown in
Fig.~\ref{fig:Cayleydist}(b) contain useful physical information
themselves. For instance, the chirality-chirality correlation
function $\langle \chi_{k} \chi_{k^\prime(k)} \rangle$ is exactly given by
summing over all the bins $d$ and counting the weight $p(d)$ with
alternating signs
\begin{equation}
\label{eqn:cayley:chirality}
\langle \chi_{k} \chi_{k^\prime(k)} \rangle = \sum_d (-1)^d p_d\,,
\end{equation}
reflecting the sign change of the chirality upon crossing a domain
wall. In our case, since $\langle |\chi | \rangle\approx 0.0005$, this
correlation function is of the order of $10^{-7}$ which turns out to
be difficult to resolve using \eqref{eqn:cayley:chirality} as this
would require an accuracy of better than $10^{-8}$ of each
histogram entry. It is, however easy to check that
Eq.~\eqref{eqn:cayley:chirality} averages to a number very close to $0$
with a large error bar that includes the true finite chirality.

The correlation function of the pseudospins at sites $k$ and $k'(k)$
belonging to the same sublattice, $\langle \sigma_{k} \sigma_{k^\prime(k)}
\rangle$, can be expressed in terms of $p_d$ as
\begin{align}
\label{eqn:cayley:spinspin}
\nonumber
\langle \sigma_{k} \sigma_{k^\prime(k)} \rangle  =& p_0 + \frac{1}{3} p_1 - \frac{2}{6}p_2 -
\frac{4}{12}p_3 + 0 p_4 + \frac{8}{48} p_5 + \frac{8}{96} p_6  \\
&\quad- \frac{8}{192} p_7 -\frac{24}{384} p_8 - \frac{8}{768} p_9 +
\frac{40}{1536} p_{10} + \cdots\,,
\end{align}
where we assumed that the distribution on the Cayley tree is isotropic,
i.e.~that any possible irreducible sequence of $d$ domain walls occurs
with equal probability.  Under this assumption, the coefficient of $p_d$ in 
Eq.~\eqref{eqn:cayley:spinspin} is given by the average value of $\langle \sigma_{k} \sigma_{k^\prime(k)}
\rangle$ at distance $d$ on the Cayley tree. In reality, the distribution on the
Cayley tree is not exactly isotropic in finite systems. However, it can be expected to become isotropic in the thermodynamic limit. For $d=2$ and $L=5760$ we have checked that 
the substitution of the distribution function shown in Fig.~\ref{fig:Cayleydist}(b) into Eq. \ref{eqn:cayley:spinspin} (based on the assumption that the distribution function is isotropic) leads to $\langle
\sigma_{k} \sigma_{k^\prime(k)} \rangle = 0.085(3)$, which indeed is nicely
consistent with the value of $m_\alpha^2$ ($0.2925^2 \simeq 0.0856$) following from the more precise estimate of the sublattice magnetization [see Eq.~\eqref{eqn:result:ms}].

\subsection{Determination of the ground-state degeneracy\label{sec:entropy}}

To round up our numeric characterization of the ground-state of the
non-coplanar phase of the classical bilinear-biquadratic Heisenberg model,
we turn to the calculation of the residual entropy per site 
characterizing the ground-state degeneracy. When a system allows for
independent local transformations that do not change its energy, the number of
ground states $\Omega(L)$ can be expected to grow exponentially with the number of sites, that is, as
\begin{equation}
\label{eqn:scalingOmega} \Omega(L)=c e^{\alpha L^2}\,.
\end{equation}
Accordingly, the entropy per site
\begin{equation}
h(L)=\frac{1}{L^2}\ln \Omega(L)\,
\end{equation}
is expected to have a finite limit at $L\to\infty$,
\begin{equation}
    \lim_{L\to\infty} h(L)=\alpha>0\;.
\end{equation}
Using a comparison with exactly solvable models Fendley \etal
\cite{Fendley2007} showed that the residual entropy $\alpha$ has to belong to
the interval \makebox{$0.3332 < \alpha < 0.3791$} and confirmed this by a
numerical analysis, which gave an estimate $\alpha\approx
0.3661$. Our intention here is to improve the accuracy of this estimate on
the basis of our efficient MC update and, in addition, to study
whether the size dependence of the ground-state degeneracy also manifests
some crossover phenomenon, in analogy to the magnetization and chirality.

Although the calculation of $\Omega$ is not directly possible within any
of the above MC methods, one can combine these methods with the
Wang-Landau scheme \cite{WL} to obtain an estimate of the density of
states $g(m)$ for all possible values of the total magnetization $m=\sum_i
\sigma_i$. A basic version of this approach consists in using only the
local MC updates in which the unnormalized density of states $\tilde g$ [which initially is assumed to be independent of $m$,  $\tilde g(m)\equiv 1$] is
multiplied by a factor $f>1$, $\tilde g(m)\to f \tilde
g(m)$, for any configuration of magnetization $m$ generated. Then, the only
difference to the procedure applied above for the calculation of averages
of different quantities consists in using as the acceptance probability
for flipping a flippable spin $\sigma_i$ the dynamic quantity $\tilde
g(m)/\tilde g(m - 2\sigma_i)$ instead of $1/2$. Conventionally, one
iteration ends once all possible magnetizations have been seen (this can
be checked by counting how often the simulation has observed
magnetizations $\pm L^2$ and if it has travelled between those extremes)
and if the histogram of the visited magnetizations is sufficiently flat.
In the first iteration one takes $f=e=2.718\ldots$, but after each iteration, the modification factor $f$ is diminished ($f \to
\sqrt{f}$) until $f$ becomes equal to $1$ within some precision $\epsilon$
(say $\epsilon=10^{-7}$).

In order to achieve the convergence at essentially larger system sizes than
those which can be studied in the framework of this canonical Wang-Landau
implementation, we modified the above basic strategy in the following (not
necessarily unique) way. First, we employ the approach proposed by Zhou
and Bhatt,\cite{Bhatt_WL_2005} which most notably relaxes the
flat-histogram criterion by defining that a Wang-Landau iteration ends
once each accessible microstate is sampled at least $1/\ln{f}$ times.
Second, we add occasional cluster moves to the sampling procedure applying
the same acceptance rules as for the local Wang-Landau updates. The
cluster updates are clearly useful as they can help to overcome barriers
and to perform large jumps to phase space regions which were not yet
 sufficiently sampled. Third -- and probably most importantly -- we
parallelized the whole procedure on a multi-core shared-memory
architecture using the concurrent approach of Ref.~\onlinecite{zhan2008}.

The density of states $\tilde g(m)$ thus obtained can be normalized using
the condition that there are only two regular states in the pseudospin
model having total magnetization $m=\pm L^2$, or $g(L^2)=g(-L^2)=1$.
Hence, an estimate for the number of ground-state configurations will be
given by
\begin{equation}
\Omega = \frac{1}{\tilde g(L^2)} \sum_{m=-L^2}^{m=L^2} \tilde g(m)\,,
\end{equation}
where an uncertainty estimate is obtained by repeating the calculation
$20$ times with different pseudo-random numbers.

Using this strategy, we obtained rather accurate estimates for the number
of ground-states $\Omega(L)$ on systems with linear size up to $L=1152$.
A finite-size analysis of these numbers reveals that the dependence
(\ref{eqn:scalingOmega}) is very nicely satisfied at all length scales,
with almost no change in behavior between small and large length scales.

\begin{figure}
\centering
\includegraphics[width=0.9\columnwidth]{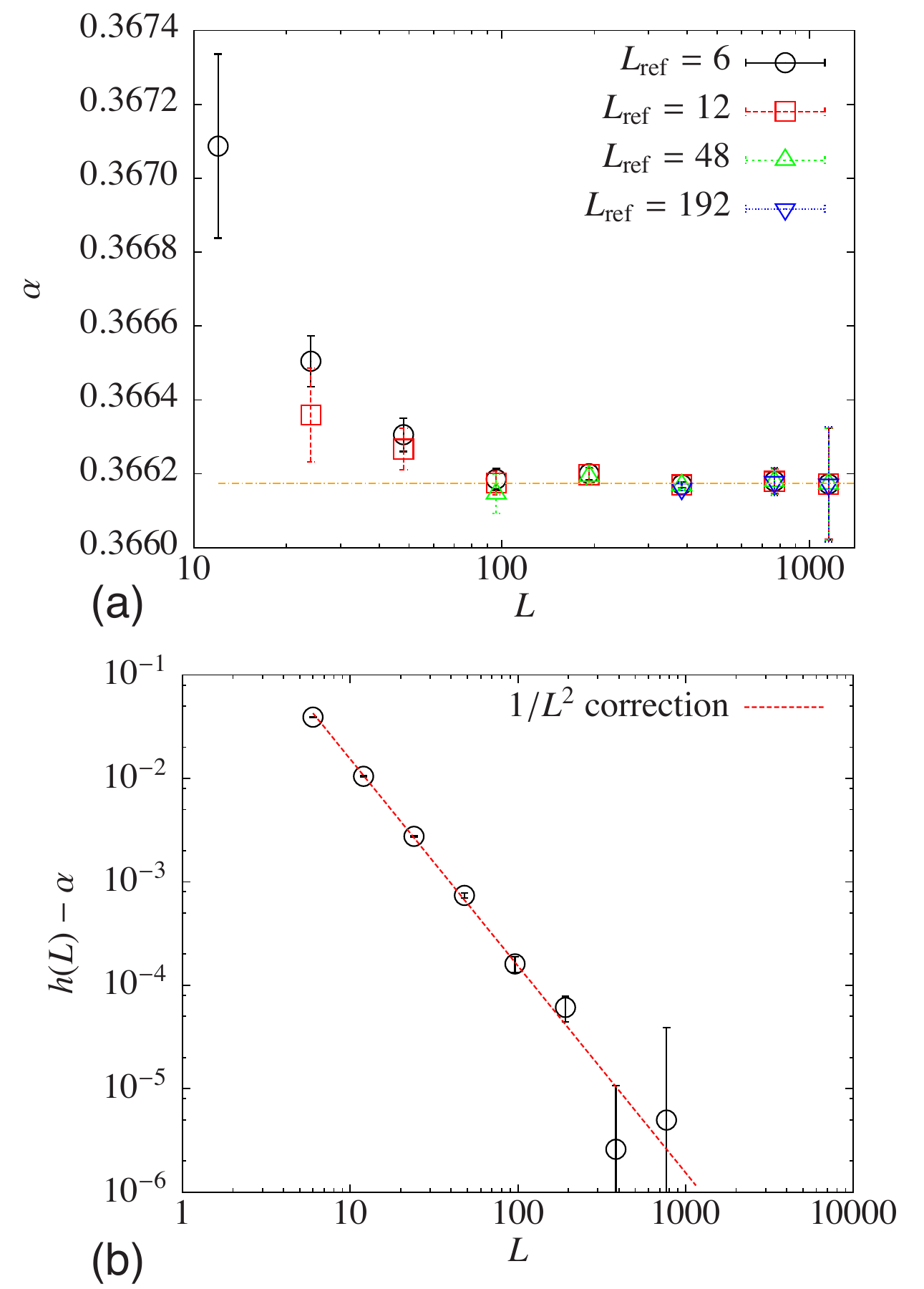}
\caption{\label{fig:improvedWL}(a) Size dependence of the exponent
  $\alpha$
  obtained from a pairwise comparison of ground-state degeneracies at
  sizes $L$ and $L_{\rm{ref}}$.  Curves using different fixed $L_{\rm ref}$
  collapse onto each other. The straight line is the result from the
  standard fit of $\Omega(L)$ to Eq. (\ref{eqn:scalingOmega}) for
  $L>24$. (b) The behavior of the size-dependent correction to entropy
  per site $h(L)$. Straight line corresponds to the $1/L^2$ dependence.}
\end{figure}

This is demonstrated in Fig.~\ref{fig:improvedWL}(a), which displays the
size dependence of the exponent $\alpha$ calculated from the ground-state
degeneracies $\Omega(L)$ and $\Omega(L_{\rm{ref}})$ at two different
system sizes $L$ and $L_{\rm{ref}}$ by using the relation
\begin{equation} \alpha = \frac{\ln\left[ \Omega(L)\right]-\ln\left[\Omega(L_{\rm{ref}})
\right]}{ \left. L^2 - L_{\rm{ref}}^2 \right.}\,.
\end{equation}
The reason to use two values of $\Omega$ at different $L$ is that this
allows one to eliminate the unknown factor $c$ in Eq.
(\ref{eqn:scalingOmega}). Curves using different choices for $L_{\rm ref}$ converge
and collapse onto each other for $L>100$, showing the
validity of the scaling formula. The finite-size deviations at small $L$
are in the $1\%$ region.

Moreover, the size dependence of the entropy per site $h(L)$ which we observe
is in perfect agreement with the predictions of the conformal
analysis\cite{Bloete86,Affleck86, Fendley2007} according to
which at $L\to \infty$ the entropy per site $h(L)$ approaches its limit
$\alpha$ as
\begin{equation}
\label{eqn:Lsquaredcorrection}
    h(L)\approx \alpha+\frac{a}{L^2} ,
\end{equation}
consistent with Eq.~(\ref{eqn:scalingOmega}). Our data adheres excellently to this scaling
assumption and allows us to obtain an improved estimate for the residual
entropy per site as
\begin{equation}\label{eqn:alpha}
\alpha=0.36618(1)\,.
\end{equation}
The data in Fig.~\ref{fig:improvedWL}(b) shows that the finite-size
corrections to this number follow the $1/L^2$ dependence represented by
the straight line.

Note that these data show no sign of a crossover effect for the scaling behavior of $\Omega(L)$ up to $L=1152$. An analogous analysis applied to the data on
the size dependence of the chirality [presented in
Fig.~\ref{fig:ClusterResults1}(a)] shows a visible tendency towards a change of regime -- from an algebraic decay to a saturation -- already at
smaller sizes, whereas in the data for the size dependence of the
magnetization [see Fig. 9(b)] the tendency to the change of regime is
visible at even smaller scales without a special analysis.
However, it has to be emphasized that in the case of $\Omega(L)$ a change of regime cannot be as drastic as for the magnetization or chirality
and can consist only in a small change of the exponent $\alpha$ and
therefore it may be much harder to observe its precursors at $L\lesssim
1000$.

The data presented in this section have been obtained by the method involving  cluster updates and therefore corresponds to counting all allowed states of the pseudospin model. We also have verified that exclusion of the states not belonging to the Heisenberg sector does not lead to any noticeable change of the results. This means that the graphs presented in Fig.~\ref{fig:improvedWL} and the value of the residual entropy given by Eq.~(\ref{eqn:alpha}) describe as well the properties of the original Heisenberg model.

\section{\label{sec:conclusions}Summary}

In conclusion, we have studied the zero-temperature properties of the non-coplanar phase of the classical bilinear-biquadratic Heisenberg model on the triangular lattice by means of extensive Monte Carlo simulations.
Building on the characterization of the non-coplanar ground states in
terms of zero-energy domain walls separating domains with different  three-sublattice states, we have introduced an Ising pseudospin representation that allows for a particularly convenient description of the ground states
without any reference to details such as the angle $\Phi$ between neighboring Heisenberg spins. The Ising representation gave rise to an elegant and
efficient Monte Carlo procedure in which the space of allowed
configurations is sampled by discrete pseudospin flips.

Both local and cluster Monte Carlo updates have been introduced, the
latter being motivated by the need to simulate unusually large system
sizes. Using a cluster Monte Carlo procedure, we have been able to
study ground state configurations in systems containing up to  $1.3\cdot 10^8$ spins.  In addition to the basic characterization of the ground state
degeneracy and of the domain-wall statistics, we have been able to
unambiguously conclude on the nature of ordering in the non-coplanar phase at zero temperature.

In terms of the pseudospin variables, evidence in favor of the presence of long-range magnetic order has been found already in systems of intermediate sizes, with linear size $L$ up to about $200$.
The existence of such an ordering had been predicted from analytical arguments in Ref. \onlinecite{Korshunov2012}. Note that the earlier work of Fendley {\em et al.} \cite{Fendley2007} was not conclusive in that respect and could not exclude a disordered state.

However, to reach a full understanding of the nature of ordering (in particular, to establish whether all three sublattices are ordered to an equal degree) turned out to be a highly nontrivial task. Only simulations on very large systems of linear size $L\gsim 1000$ have given evidence that in the thermodynamic limit all three sublattices become equivalent, whereas below this length scale numerical simulations have not allowed the exclusion of a partially disordered state.

Although the existence of long-range order for Heisenberg spins follows already from the existence of the long-range order in the pseudospin model, \cite{Korshunov2012}
we have also provided an additional proof for that directly for the Heisenberg spins by analyzing the distribution of the system in its order parameter space, which has the structure of an infinite Cayley tree of degree 3.
It was shown that this distribution is highly localized and that
two spin domains which are infinitely distant from each other
are only separated on average by a finite number (roughly $2$) of irreducible domain walls. This ensures the presence of long-range correlations for Heisenberg spins.

Importantly, the large-length simulations also provided evidence in favor of long-range chiral correlations in the pseudospin model, as well as for the
non-coplanar ground states of the original Heisenberg model. In contrast to the ordering related to the orientation of Heisenberg spins (which at an arbitrarily low temperature is destroyed by continuous fluctuations - spin waves\cite{Polyakov,Azaria92}), this order, which is related to a discrete degree of freedom, can be expected to survive at low enough temperatures, leading to a finite-temperature phase transition. The results of simulations of the bilinear-biquadratic Heisenberg model on the triangular lattice at finite temperatures have been reported in Ref. \onlinecite{Kawamura2007}. However, the system sizes in this work never exceeded two hundred, and therefore the observation of ordering at low temperatures was rather a finite-size effect than the consequence of real chiral ordering which, according to our findings, is likely to exist only at much larger length scales.

We have also performed a high precision numerical calculation of the residual entropy of the model on the basis of a generalization of the Wang-Landau scheme. 
Our results agree with those of Ref. \onlinecite{Fendley2007}, but the accuracy has been increased by one order of magnitude. To our surprise, the size dependence of the ground state degeneracy (which determines the value of the residual entropy) has not shown any sign of a crossover at $L\sim 1000$, as could have been expected from the existence of such a crossover in the behavior of other quantities (magnetization and chirality).

The most unusual feature of the model is the existence of a large length scale below which long-range order in terms of magnetization is combined with an {\em algebraic} decay of fluctuations and the complete disorder of the chirality. Only at sizes $L\gsim 1000$ does the behavior become more consistent with long-range order both in terms of the magnetization and chirality, and with an exponential leveling off of correlations.

The long-range order of the magnetization and of the chirality disappears in the special case of $\Phi=\arccos(-1/3)$, when the model becomes equivalent to the exactly solvable \cite{Baxter70} four-state antiferromagnetic Potts model on the triangular lattice, which is known to be critical. In terms of the loop representation this case corresponds to the presence of one more allowed vertex \cite{Fendley2007,Korshunov2012}   (a merging point of {\em six} domain walls) which destroys the possibility of constructing the pseudospin description of loop configurations. It seems plausible that even when they are allowed, such complex objects can be present only in low concentrations, because their presence strongly suppresses the possibilities of domain wall fluctuations. However, it eludes our understanding why the appearance of such rare defects shifts the system exactly into the critical point. In more familiar situations, the presence of rare topological defects induces the appearance of a finite correlation radius related to their concentration (like it happens in the antiferromagnetic Ising model on the triangular lattice at a finite temperature \cite{Nienhuis84} or in the conventional XY model above the Berezinkii-Kosterlitz-Thouless phase transition) and not its growth to infinity. 
In fact, the extreme largeness of the correlation radii we observed may precisely be related to the smallness of the concentration of the defects whose presence distinguishes the generic model from the special case $\Phi=\arccos(-1/3)$ situated exactly at criticality. In our opinion, this point deserves further investigation.

It should also be mentioned that the existence of a large length scale at which the properties of the system change qualitatively raises some
questions about the analysis of Ref. \onlinecite{Fendley2007}, whose aim was to extract the central charge from the size dependence of the correction to the residual entropy. Our findings suggest that, since the numerical analysis of Ref. \onlinecite{Fendley2007} relies on cylindrical systems with transverse sizes only up to $L=36$, it cannot be expected to provide reliable information on the behavior of the system in the thermodynamic limit.

\acknowledgements
We are grateful for the support of the Swiss National Science Foundation, MaNEP, and the Hungarian OTKA Grant No. K106047 and are indebted to M.~Hasenbusch and T.~Vogel for useful discussions.

\bibliographystyle{apsrev4-1}
\bibliography{literature}

\begin{thebibliography}{26}
\expandafter\ifx\csname natexlab\endcsname\relax\def\natexlab#1{#1}\fi
\expandafter\ifx\csname bibnamefont\endcsname\relax
  \def\bibnamefont#1{#1}\fi
\expandafter\ifx\csname bibfnamefont\endcsname\relax
  \def\bibfnamefont#1{#1}\fi
\expandafter\ifx\csname citenamefont\endcsname\relax
  \def\citenamefont#1{#1}\fi
\expandafter\ifx\csname url\endcsname\relax
  \def\url#1{\texttt{#1}}\fi
\expandafter\ifx\csname urlprefix\endcsname\relax\def\urlprefix{URL }\fi
\providecommand{\bibinfo}[2]{#2}
\providecommand{\eprint}[2][]{\url{#2}}

\bibitem[{\citenamefont{Sutherland}(1975)}]{sutherland}
\bibinfo{author}{\bibfnamefont{B.}~\bibnamefont{Sutherland}},
  \bibinfo{journal}{Phys. Rev. B} \textbf{\bibinfo{volume}{12}},
  \bibinfo{pages}{3795} (\bibinfo{year}{1975}).

\bibitem[{\citenamefont{{Takhtajan}}(1982)}]{takhtajan}
\bibinfo{author}{\bibfnamefont{L.~A.} \bibnamefont{{Takhtajan}}},
  \bibinfo{journal}{Physics Letters A} \textbf{\bibinfo{volume}{87}},
  \bibinfo{pages}{479} (\bibinfo{year}{1982}).

\bibitem[{\citenamefont{{Babujian}}(1982)}]{babujian}
\bibinfo{author}{\bibfnamefont{H.~M.} \bibnamefont{{Babujian}}},
  \bibinfo{journal}{Physics Letters A} \textbf{\bibinfo{volume}{90}},
  \bibinfo{pages}{479} (\bibinfo{year}{1982}).

\bibitem[{\citenamefont{Affleck et~al.}(1987)\citenamefont{Affleck, Kennedy,
  Lieb, and Tasaki}}]{AKLT}
\bibinfo{author}{\bibfnamefont{I.}~\bibnamefont{Affleck}},
  \bibinfo{author}{\bibfnamefont{T.}~\bibnamefont{Kennedy}},
  \bibinfo{author}{\bibfnamefont{E.~H.} \bibnamefont{Lieb}}, \bibnamefont{and}
  \bibinfo{author}{\bibfnamefont{H.}~\bibnamefont{Tasaki}},
  \bibinfo{journal}{Phys. Rev. Lett.} \textbf{\bibinfo{volume}{59}},
  \bibinfo{pages}{799} (\bibinfo{year}{1987}).

\bibitem[{\citenamefont{{Nakatsuji} et~al.}(2005)\citenamefont{{Nakatsuji},
  {Nambu}, {Tonomura}, {Sakai}, {Jonas}, {Broholm}, {Tsunetsugu}, {Qiu}, and
  {Maeno}}}]{Nakatsuji2005}
\bibinfo{author}{\bibfnamefont{S.}~\bibnamefont{{Nakatsuji}}},
  \bibinfo{author}{\bibfnamefont{Y.}~\bibnamefont{{Nambu}}},
  \bibinfo{author}{\bibfnamefont{H.}~\bibnamefont{{Tonomura}}},
  \bibinfo{author}{\bibfnamefont{O.}~\bibnamefont{{Sakai}}},
  \bibinfo{author}{\bibfnamefont{S.}~\bibnamefont{{Jonas}}},
  \bibinfo{author}{\bibfnamefont{C.}~\bibnamefont{{Broholm}}},
  \bibinfo{author}{\bibfnamefont{H.}~\bibnamefont{{Tsunetsugu}}},
  \bibinfo{author}{\bibfnamefont{Y.}~\bibnamefont{{Qiu}}}, \bibnamefont{and}
  \bibinfo{author}{\bibfnamefont{Y.}~\bibnamefont{{Maeno}}},
  \bibinfo{journal}{Science} \textbf{\bibinfo{volume}{309}},
  \bibinfo{pages}{1697} (\bibinfo{year}{2005}).

\bibitem[{\citenamefont{Harada and Kawashima}(2002)}]{Harada2002}
\bibinfo{author}{\bibfnamefont{K.}~\bibnamefont{Harada}} \bibnamefont{and}
  \bibinfo{author}{\bibfnamefont{N.}~\bibnamefont{Kawashima}},
  \bibinfo{journal}{Phys. Rev. B} \textbf{\bibinfo{volume}{65}},
  \bibinfo{pages}{052403} (\bibinfo{year}{2002}).

\bibitem[{\citenamefont{Tsunetsugu and Arikawa}(2006)}]{Tsunetsugu2006}
\bibinfo{author}{\bibfnamefont{H.}~\bibnamefont{Tsunetsugu}} \bibnamefont{and}
  \bibinfo{author}{\bibfnamefont{M.}~\bibnamefont{Arikawa}},
  \bibinfo{journal}{Journal of the Physical Society of Japan}
  \textbf{\bibinfo{volume}{75}}, \bibinfo{pages}{083701}
  (\bibinfo{year}{2006}).

\bibitem[{\citenamefont{Bhattacharjee et~al.}(2006)\citenamefont{Bhattacharjee,
  Shenoy, and Senthil}}]{Bhattarchajee2006}
\bibinfo{author}{\bibfnamefont{S.}~\bibnamefont{Bhattacharjee}},
  \bibinfo{author}{\bibfnamefont{V.~B.} \bibnamefont{Shenoy}},
  \bibnamefont{and} \bibinfo{author}{\bibfnamefont{T.}~\bibnamefont{Senthil}},
  \bibinfo{journal}{Phys. Rev. B} \textbf{\bibinfo{volume}{74}},
  \bibinfo{pages}{092406} (\bibinfo{year}{2006}).

\bibitem[{\citenamefont{L\"auchli et~al.}(2006)\citenamefont{L\"auchli, Mila,
  and Penc}}]{Laeuchli2006}
\bibinfo{author}{\bibfnamefont{A.}~\bibnamefont{L\"auchli}},
  \bibinfo{author}{\bibfnamefont{F.}~\bibnamefont{Mila}}, \bibnamefont{and}
  \bibinfo{author}{\bibfnamefont{K.}~\bibnamefont{Penc}},
  \bibinfo{journal}{Phys. Rev. Lett.} \textbf{\bibinfo{volume}{97}},
  \bibinfo{pages}{087205} (\bibinfo{year}{2006}).

\bibitem[{\citenamefont{T\'oth et~al.}(2012)\citenamefont{T\'oth, L\"auchli,
  Mila, and Penc}}]{Toth2012}
\bibinfo{author}{\bibfnamefont{T.~A.} \bibnamefont{T\'oth}},
  \bibinfo{author}{\bibfnamefont{A.~M.} \bibnamefont{L\"auchli}},
  \bibinfo{author}{\bibfnamefont{F.}~\bibnamefont{Mila}}, \bibnamefont{and}
  \bibinfo{author}{\bibfnamefont{K.}~\bibnamefont{Penc}},
  \bibinfo{journal}{Phys. Rev. B} \textbf{\bibinfo{volume}{85}},
  \bibinfo{pages}{140403} (\bibinfo{year}{2012}).

\bibitem[{\citenamefont{Kawamura and Yamamoto}(2007)}]{Kawamura2007}
\bibinfo{author}{\bibfnamefont{H.}~\bibnamefont{Kawamura}} \bibnamefont{and}
  \bibinfo{author}{\bibfnamefont{A.}~\bibnamefont{Yamamoto}},
  \bibinfo{journal}{Journal of the Physical Society of Japan}
  \textbf{\bibinfo{volume}{76}}, \bibinfo{pages}{073704}
  (\bibinfo{year}{2007}).

\bibitem[{\citenamefont{{Korshunov} et~al.}(2012)\citenamefont{{Korshunov},
  {Mila}, and {Penc}}}]{Korshunov2012}
\bibinfo{author}{\bibfnamefont{S.~E.} \bibnamefont{{Korshunov}}},
  \bibinfo{author}{\bibfnamefont{F.}~\bibnamefont{{Mila}}}, \bibnamefont{and}
  \bibinfo{author}{\bibfnamefont{K.}~\bibnamefont{{Penc}}},
  \bibinfo{journal}{Phys. Rev. B} \textbf{\bibinfo{volume}{85}},
  \bibinfo{pages}{174420} (\bibinfo{year}{2012}).

\bibitem[{\citenamefont{{Fendley} et~al.}(2007)\citenamefont{{Fendley},
  {Moore}, and {Xu}}}]{Fendley2007}
\bibinfo{author}{\bibfnamefont{P.}~\bibnamefont{{Fendley}}},
  \bibinfo{author}{\bibfnamefont{J.~E.} \bibnamefont{{Moore}}},
  \bibnamefont{and} \bibinfo{author}{\bibfnamefont{C.}~\bibnamefont{{Xu}}},
  \bibinfo{journal}{Phys. Rev. E} \textbf{\bibinfo{volume}{75}},
  \bibinfo{eid}{051120} (\bibinfo{year}{2007}).

\bibitem[{\citenamefont{Wolff}(1989)}]{Wolff89}
\bibinfo{author}{\bibfnamefont{U.}~\bibnamefont{Wolff}},
  \bibinfo{journal}{Phys. Rev. Lett.} \textbf{\bibinfo{volume}{62}},
  \bibinfo{pages}{361} (\bibinfo{year}{1989}).

\bibitem[{\citenamefont{Swendsen and Wang}(1987)}]{SwendsenWang97}
\bibinfo{author}{\bibfnamefont{R.~H.} \bibnamefont{Swendsen}} \bibnamefont{and}
  \bibinfo{author}{\bibfnamefont{J.-S.} \bibnamefont{Wang}},
  \bibinfo{journal}{Phys. Rev. Lett.} \textbf{\bibinfo{volume}{58}},
  \bibinfo{pages}{86} (\bibinfo{year}{1987}).

\bibitem[{\citenamefont{{Mart{\'{\i}}n-Herrero}}(2004)}]{Herrero2004JPhA}
\bibinfo{author}{\bibfnamefont{J.}~\bibnamefont{{Mart{\'{\i}}n-Herrero}}},
  \bibinfo{journal}{Journal of Physics A Mathematical General}
  \textbf{\bibinfo{volume}{37}}, \bibinfo{pages}{9377} (\bibinfo{year}{2004}).

\bibitem[{\citenamefont{Press et~al.}(2007)\citenamefont{Press, Teukolsky,
  Vetterling, and Flannery}}]{press2007numerical}
\bibinfo{author}{\bibfnamefont{W.}~\bibnamefont{Press}},
  \bibinfo{author}{\bibfnamefont{S.}~\bibnamefont{Teukolsky}},
  \bibinfo{author}{\bibfnamefont{W.}~\bibnamefont{Vetterling}},
  \bibnamefont{and} \bibinfo{author}{\bibfnamefont{B.}~\bibnamefont{Flannery}},
  \emph{\bibinfo{title}{Numerical Recipes 3rd Edition: The Art of Scientific
  Computing}} (\bibinfo{publisher}{Cambridge University Press},
  \bibinfo{year}{2007}).

\bibitem[{\citenamefont{Wang and Landau}(2001)}]{WL}
\bibinfo{author}{\bibfnamefont{F.}~\bibnamefont{Wang}} \bibnamefont{and}
  \bibinfo{author}{\bibfnamefont{D.~P.} \bibnamefont{Landau}},
  \bibinfo{journal}{Phys. Rev. Lett.} \textbf{\bibinfo{volume}{86}},
  \bibinfo{pages}{2050} (\bibinfo{year}{2001}).

\bibitem[{\citenamefont{Zhou and Bhatt}(2005)}]{Bhatt_WL_2005}
\bibinfo{author}{\bibfnamefont{C.}~\bibnamefont{Zhou}} \bibnamefont{and}
  \bibinfo{author}{\bibfnamefont{R.~N.} \bibnamefont{Bhatt}},
  \bibinfo{journal}{Phys. Rev. E} \textbf{\bibinfo{volume}{72}},
  \bibinfo{pages}{025701} (\bibinfo{year}{2005}).

\bibitem[{\citenamefont{Zhan}(2008)}]{zhan2008}
\bibinfo{author}{\bibfnamefont{L.}~\bibnamefont{Zhan}}, \bibinfo{journal}{Comp.
  Phys. Comm.} \textbf{\bibinfo{volume}{179}}, \bibinfo{pages}{339}
  (\bibinfo{year}{2008}).

\bibitem[{\citenamefont{Bl\"ote et~al.}(1986)\citenamefont{Bl\"ote, Cardy, and
  Nightingale}}]{Bloete86}
\bibinfo{author}{\bibfnamefont{H.~W.~J.} \bibnamefont{Bl\"ote}},
  \bibinfo{author}{\bibfnamefont{J.~L.} \bibnamefont{Cardy}}, \bibnamefont{and}
  \bibinfo{author}{\bibfnamefont{M.~P.} \bibnamefont{Nightingale}},
  \bibinfo{journal}{Phys. Rev. Lett.} \textbf{\bibinfo{volume}{56}},
  \bibinfo{pages}{742} (\bibinfo{year}{1986}).

\bibitem[{\citenamefont{Affleck}(1986)}]{Affleck86}
\bibinfo{author}{\bibfnamefont{I.}~\bibnamefont{Affleck}},
  \bibinfo{journal}{Phys. Rev. Lett.} \textbf{\bibinfo{volume}{56}},
  \bibinfo{pages}{746} (\bibinfo{year}{1986}).

\bibitem[{\citenamefont{Polyakov}(1975)}]{Polyakov}
\bibinfo{author}{\bibfnamefont{A.~M.} \bibnamefont{Polyakov}},
  \bibinfo{journal}{Phys. Lett. B} \textbf{\bibinfo{volume}{59}},
  \bibinfo{pages}{79} (\bibinfo{year}{1975}), ISSN \bibinfo{issn}{0370-2693}.

\bibitem[{\citenamefont{Azaria et~al.}(1992)\citenamefont{Azaria, Delamotte,
  Jolicoeur, and Mouhanna}}]{Azaria92}
\bibinfo{author}{\bibfnamefont{P.}~\bibnamefont{Azaria}},
  \bibinfo{author}{\bibfnamefont{B.}~\bibnamefont{Delamotte}},
  \bibinfo{author}{\bibfnamefont{T.}~\bibnamefont{Jolicoeur}},
  \bibnamefont{and} \bibinfo{author}{\bibfnamefont{D.}~\bibnamefont{Mouhanna}},
  \bibinfo{journal}{Phys. Rev. B} \textbf{\bibinfo{volume}{45}},
  \bibinfo{pages}{12612} (\bibinfo{year}{1992}).

\bibitem[{\citenamefont{Baxter}(1970)}]{Baxter70}
\bibinfo{author}{\bibfnamefont{R.~J.} \bibnamefont{Baxter}},
  \bibinfo{journal}{J. Math. Phys.} \textbf{\bibinfo{volume}{11}},
  \bibinfo{pages}{784} (\bibinfo{year}{1970}).

\bibitem[{\citenamefont{Nienhuis et~al.}(1984)\citenamefont{Nienhuis, Hilhorst,
  and Blote}}]{Nienhuis84}
\bibinfo{author}{\bibfnamefont{B.}~\bibnamefont{Nienhuis}},
  \bibinfo{author}{\bibfnamefont{H.~J.} \bibnamefont{Hilhorst}},
  \bibnamefont{and} \bibinfo{author}{\bibfnamefont{H.~W.~J.}
  \bibnamefont{Blote}}, \bibinfo{journal}{J. Phys. A}
  \textbf{\bibinfo{volume}{17}}, \bibinfo{pages}{3559} (\bibinfo{year}{1984}).

\end{thebibliography}

\end{document}